\documentclass[12pt,epsf]{article}

\usepackage{graphicx,float}
\usepackage{mathrsfs,array,multirow}
\usepackage{amssymb,amsfonts}
\usepackage{amsmath,amstext}
\usepackage{color}
\usepackage{subfigure}

\newcommand{\be}{\begin{equation}}

\newcommand{\ee}{\end{equation}}

\newcommand{\ba}{\begin{array}}

\newcommand{\ea}{\end{array}}

\begin{document}
\begin{titlepage}
\vspace{.5in}
\begin{flushright}

\end{flushright}
\vspace{0.5cm}

\begin{center}
{\Large\bf Phase transition for black holes in dilatonic Einstein-Gauss-Bonnet theory of gravitation}\\
\vspace{.4in}

  {$\rm{Sunly \,\, Khimphun}^{\S}$}\footnote{\it email:kpslourk@sogang.ac.kr}\,\,
  {$\rm{Bum-Hoon \,\, Lee}^{\P\S\dag}$}\footnote{\it email:bhl@sogang.ac.kr}\,\,
  {$\rm{Wonwoo \,\, Lee}^{\P}$}\footnote{\it email:warrior@sogang.ac.kr}\,\, \\

  {\small \P \it Center for Quantum Spacetime, Sogang University, Seoul 04107, Korea}\\
  {\small \S \it Department of Physics, Sogang University, Seoul 04107,
  Korea}\\
  {\small \dag \it Asia Pacific Center for Theoretical Physics, Pohang 37673, Korea}\\

\vspace{.5in}
\today
\end{center}

\begin{center}
{\large\bf Abstract}
\end{center}
\begin{center}
\begin{minipage}{4.75in}

{\small \,\,\,\, We study the thermodynamic properties of a black hole and the Hawking-Page phase transition in the asymptotically anti--de Sitter spacetime in the dilatonic Einstein-Gauss-Bonnet theory of gravitation. We show how the higher-order curvature terms can influence both the thermodynamic properties and the phase transition. We evaluate both heat capacity and free energy difference to determine the local and global thermodynamic stabilities, respectively. We find that the phase transition occurs from the thermal anti--de Sitter to a small spherical black hole geometry and occurs to a hyperbolic black hole geometry in the (dilatonic) Einstein-Gauss-Bonnet theory of gravitation unlike those in Einstein's theory of gravitation.  }


\end{minipage}
\end{center}
\end{titlepage}

\newpage

\section{INTRODUCTION \label{sec1}}

The black hole entropy has an extraordinary geometrical interpretation, which is proportional to the area of the horizon rather than the volume of a black hole. That is widely known as the Bekenstein-Hawking entropy \cite{beken, hawen}. The entropy can be derived from a variety of methods. This quantity can be obtained by the Noether charge of diffeomorphisms under the Killing vector field \cite{jamy, wald8}. This can be also obtained through the thermodynamic relation after evaluating the Euclidean action by the path integral approach \cite{giha}. This can be derived from the first law of black hole thermodynamics \cite{bcha}, in which the integration gives rise to the constant part of the entropy. In Einstein's theory of gravitation (Einstein theory), this constant part vanishes because the entropy goes to zero when the black hole mass vanishes. Consequently, there exists no discontinuous jump in the entropy when approaching the vanishing mass. However, the Gauss-Bonnet (GB) term in four dimensions causes the additional constant entropy. This quantity is related to the information on the topology of the spacetime manifold \cite{DeWitt:1975ys, swh, tkm3, Liko:2007vi, sawa}, which means that there exists a discontinuous jump when the topology of the manifold is changed or the black hole mass vanishes away, unlike the limiting case in Einstein theory.

A black hole in anti--de Sitter (AdS) spacetime has the minimum temperature, unlike the Schwarzschild black hole with a monotonically decreasing temperature. The decreasing part of the temperature corresponds to the negative specific heat, while the increasing part of the temperature corresponds to the positive specific heat unlike the Schwarzschild black hole with only a decreasing negative specific heat. As a consequence, the black hole is locally unstable below the radius with the minimum temperature, while the black hole above the radius can be in a stable thermodynamic equilibrium. For this reason, one can employ the canonical ensemble description.

The Hawking-Page (HP) phase transition represents the transition occurring between the AdS black holes and the thermal AdS space as a first-order phase transition. The description of the phase transition is based on the free energy difference obtained by using the Euclidean path integral approach for the black hole thermodynamics, in which the free energy vanishes for the thermal AdS space with the spherical symmetry \cite{HawPa}. According to the analysis of the free energy and the specific heat, there exists only the thermal AdS space below the minimum temperature. Above the minimum temperature, there exist two types of black holes depending on the sign of a specific heat. On the other hand, there also exist two types of black holes depending on the sign of a free energy difference. One is in a globally unstable phase with the positive free energy difference, in which the thermal AdS space is more probable than the AdS black hole. The other is in a globally stable phase with the negative free energy difference, in which the black hole is more probable than the thermal AdS space. The phase transition occurs at the crossing point with the vanishing free energy difference. Consequently, there are three types of black holes in AdS space depending on two instabilities. One belongs to both locally and globally unstable phases, another to locally stable and globally unstable phases, and the other to both locally and globally stable phases. These are definitively divided into three types in terms of the radius of a black hole. The HP phase transition has been extensively studied in Refs.\ \cite{ysm10, ekyi, mymo, giki, ysm11, egki, dema, maha} and with curvature terms in higher dimensions \cite{rgc0, chne, neup, cakw, mkp}.

If we think over the very early universe approaching the Planck scale, we could consider that the quantum effect can become more and more important even if we do not know the perfect quantum theory of gravitation. One way to realize this theory is to adopt the theory of gravitation with higher-order curvature terms as an effective theory to probe the effect of quantum gravity at the low energy \cite{campf, Zwiebach, boude, anrt, kllt, agll}. Unfortunately, such a theory suffers from the unnecessary ghost and unitary problems. However, the theory with GB term does not have those problems. Furthermore, the GB term can be considered as a topological invariant quantity in four dimensions, that is, the action integral with the GB term provides the information on the topology of a spacetime manifold. Because the contribution from the GB term vanishes under variation in four dimensions, the existence of the term does not influence both the equations of motion and the solutions. How can we distinguish the difference between Einstein theory and the Einstein-Gauss-Bonnet theory of gravitation (EGB theory) in four dimensions? To answer this question is one of the purposes of this investigation. In order to consider the effect of the GB term on both the background spacetime and the field evolution, the GB term is required to be coupled to the matter field, i.e., one adopts the dilatonic Einstein-Gauss-Bonnet theory of gravitation (DEGB theory). The black hole in DEGB theory has the scalar field as a secondary hair \cite{copw, kanti01}, which has been extensively studied \cite{tyma02, kanti02, toma3, got03, got04, ohto04}.

We consider a canonical ensemble description for the asymptotically AdS black hole geometry; therefore, we employ the Euclidean path integral approach for the black hole thermodynamics. We first evaluate the Euclidean action with the higher-order curvature terms. The entropy and the free energy should be straightforwardly modified thanks to the higher-order curvature terms. With a certain sign in front of the terms, the free energy could decrease, whereas the entropy could increase \cite{jamy, tkm3, tkm4}. The counterterms to renormalize the infinite Euclidean action in AdS space have been studied for the various contexts \cite{bakr, ejmy, lisa, park01, park02}.

The organization of this paper is as follows: In the next section, we set up the basic framework for this paper. We numerically construct the hairy black holes by solving the equations of motion in the asymptotically AdS spacetime action according to Refs.\ \cite{got04, ohto04}. We consider both spherically symmetric solutions with $k = 1$ and hyperbolic solutions with $k = -1$. In Sec.\ \ref{sec3}, we study the black hole thermodynamic properties and the HP phase transition in (D)EGB theories, which has not been studied before. We employ the Euclidean path integral formalism to straightforwardly evaluate the Euclidean action, entropy and free energy for the EGB black hole. Since one cannot obtain the analytic calculation of the Euclidean action as well as a counterterm for the black hole in DEGB theory, we adopt the entropy formula obtained by the different method. Using the thermodynamic relation we obtain the free energy. For the analysis of the HP phase transition, we take the thermal AdS with $M = 0$ for $k = 1$ and with $M = M_{\rm crit}$ for $k =-1$ as the reference background geometry and analyze those in detail. We also numerically compute the temperature and the specific heat of the asymptotically AdS black holes. In the final section, we summarize our results and discuss possible future applications.

\section{ASYMPTOTICALLY BLACK HOLD SOLUTION WITH HIGHER-ORDER CURVATURE TERMS \label{sec2}}

We numerically construct the asymptotically AdS black hole solution in DEGB theory. The numerical solutions and properties of those black holes are presented.

\subsection{The model }

We consider the action
\begin{equation}
I=  \int_{\mathcal M} \sqrt{-g} d^4 x \left[ \frac{R+2\Lambda e^{\lambda \Phi}}{2\kappa}
-\frac{1}{2}{\nabla^\mu}\Phi {\nabla_\mu}\Phi +\alpha e^{-\gamma\Phi}R^2_{\rm GB} \right]
- \oint_{\partial \mathcal M} \sqrt{-h} d^3 x \frac{K}{\kappa} + I_{\rm GBb} + I_{\rm ct}\,,
\label{f-action}
\end{equation}
where $g=\det g_{\mu\nu}\,$, $\kappa \equiv 8\pi G\,$, and $R\,$ denotes the scalar curvature of the spacetime $\mathcal M$\,. The negative cosmological constant ($\Lambda > 0$ in the present paper) is considered for the AdS geometry. The higher curvature GB term is given by $R^{2}_{\rm GB} = R^2 - 4 R_{\mu\nu}R^{\mu\nu} +  R_{\mu\nu\rho\sigma} R^{\mu\nu\rho\sigma}$\,. The scalar field $\Phi\,$ is coupled with the GB term by the GB coupling $\alpha e^{-\gamma\Phi}\,$. The parameters $\alpha\,$, $\gamma\,$, and $\lambda\,$ are constants. The second term on the right-hand side is the boundary term \cite{York, giha}, in which $h$ is the determinant of the first fundamental form, and $K$ is the trace of the second fundamental form of the boundary $\partial \mathcal M$ for the metric $g_{\mu\nu}$. The third term is the boundary term for the GB term \cite{Myers:1987yn, Davis:2002gn, Brihaye:2008xu}. The last term $I_{\rm ct}$ is the counterterm to renormalize the Euclidean action for the AdS geometry \cite{bakr}. We adopt the sign convention in Ref.\ \cite{misner}. In this section, we follow the procedure to be consistent with our action according to Refs.\ \cite{got04, ohto04}.

For the couplings $\gamma=0$ and $\lambda=0$, the DEGB theory is reduced to the EGB theory. For the coupling $\Lambda=0$, the theory is reduced to the system in asymptotically flat spacetime \cite{agll}. For the coupling $\alpha=0$ and $\lambda=0$, the solution goes to a Schwarzschild anti--de Sitter black hole in Einstein theory. We choose the positive $\gamma$ values, in which the action (\ref{f-action}) has a symmetry under $\gamma \rightarrow -\gamma$ and $\Phi \rightarrow -\Phi$. In this paper we keep $\alpha$ explicitly, because the different values of $\alpha$ will influence directly the black hole thermodynamics and the HP phase transition.

From the variation of the action (\ref{f-action}), we obtain
\begin{equation}
R_{\mu\nu} - \frac{1}{2}R g_{\mu\nu} - \Lambda e^{\lambda \Phi}g_{\mu\nu} = \kappa \left( \partial_{\mu}
\Phi \partial_{\nu}\Phi -\frac{1}{2}g_{\mu\nu}
\partial_{\rho} \Phi \partial^{\rho} \Phi  + T_{\mu\nu}^{GB}
\right)\,,
\end{equation}
\begin{equation}
\frac{1}{\sqrt{-g}} \partial_{\mu} [\sqrt{-g}
g^{\mu\nu}\partial_{\nu}\Phi] - \alpha\gamma e^{-\gamma\Phi} R_{GB}^2 + \frac{\Lambda\lambda e^{\lambda \Phi}}{\kappa} =0\,, \label{tmn}
\end{equation}
where the GB term contributes to the energy-momentum tensor
\begin{eqnarray}
T_{\mu\nu}^{GB} &=& 8 \alpha (-R_{\mu\rho\nu\sigma}\nabla^{\rho}\nabla^{\sigma} e^{-\gamma\Phi}
+R_{\mu\nu}\square e^{-\gamma\Phi}  - 2\nabla_{\rho} \nabla_{(\mu} e^{-\gamma\Phi} {R^{\rho}}_{\nu)}
+\frac{1}{2} R\nabla_{\mu}\nabla_{\nu} e^{-\gamma\Phi} )
\nonumber \\
&&+4\alpha(2R^{\rho\sigma}\nabla_{\rho}\nabla_{\sigma}  e^{-\gamma\Phi}
 - R\square e^{-\gamma\Phi} ) g_{\mu\nu}\,,
\end{eqnarray}
and $\square \equiv \nabla_{\mu} \nabla^{\mu}$\, is the d'Alembertian.

We consider the metric ansatz
\begin{equation}
ds^2 = - B(r)e^{-2 \delta(r)} dt^2 + B(r)^{-1} dr^2+ r^2 h_{ij}dx^i dx^j, \label{metric}
\end{equation}
where $h_{ij}dx^i dx^j$ represents the line element of two-dimensional surfaces of constant positive, zero, or negative curvature $2k$ according to whether $k=+1$, $0$, or $-1$, respectively. When $\gamma$, $\lambda$ and $\Phi$ vanish, the metric function $\delta(r)$ vanishes and $B(r)$ reduces to $k - \frac{2GM}{r^2} + \frac{\Lambda}{3} r^2$.

The field equation turns out to be
\begin{eqnarray}
&&\Phi''+\left(\frac{B'}{B} -\delta' + \frac{2}{r}  \right)\Phi'  +\frac{4\alpha\gamma e^{-\gamma\Phi}B}{r^2} \left(2\delta''B -2 \delta'^{2}B -B'' + k \frac{B''}{B} + 5\delta'B' \right.  \nonumber \\
&&\left. - \frac{B'^{2}}{B} -3k\delta'\frac{B'}{B} -2k\delta''+2k\delta'^{2} + \frac{\lambda r^2 \Lambda e^{\Phi(\lambda+\gamma)}}{4\alpha\gamma\kappa B}  \right)  =0 \,, \label{fieeq}
\end{eqnarray}
where the prime denotes the differentiation with respect to $r$. And there are three Einstein equations after some rearrangement as follows:
\begin{eqnarray}
&&2B'[4\alpha\kappa\gamma\Phi'(3B-k) + r e^{\gamma\Phi}] - \kappa  B \Phi'^{2}[16\alpha\gamma^2(B-k)-r^2e^{\gamma\Phi}]    \nonumber \\
&&+16\alpha\kappa\gamma B(B-k)\Phi'' + 2e^{\gamma\Phi}(B-k- \Lambda r^2 e^{\lambda\Phi})=0 \,, \label{gtt}
\end{eqnarray}
\begin{eqnarray}
\delta' \left[4\alpha\gamma\kappa(3B-k) e^{-\gamma\Phi}\Phi'+ r\right] +\frac{1}{2}\kappa r^2\Phi'^2 - 4\alpha\gamma\kappa(k-B) e^{-\gamma\Phi} \left(\Phi''-\gamma\Phi'^2\right)=0 \,. \label{DeltaEq}
\end{eqnarray}
\begin{eqnarray}
&&B \kappa r e^{-\gamma\Phi-\delta} \left[32\alpha^2 \gamma ^2(B-k)B'-64B\alpha^2\gamma^2\kappa (B-k)\delta'+r^2 e^{\gamma\Phi} (4B\alpha\gamma\kappa\Phi'+r e^{\gamma\Phi})\right]\Phi'' \nonumber \\
&& -e^{-\gamma  \Phi -\delta } \left[-B'\kappa \left\{8 \alpha\gamma (B-k) e^{\gamma\Phi}-8 B\alpha \gamma\kappa \Phi'^2 \left(4\alpha \gamma^2 (B-k)-r^2 e^{\gamma\Phi}\right)\right.\right. \nonumber \\
&&\left.\left. +8 B\alpha\gamma\delta' \left(8\alpha\gamma\kappa k \Phi'+r e^{\gamma\Phi}\right)+r^3 e^{2\gamma\Phi} \Phi'\right\}+4 \alpha\gamma\kappa B'^2 \left(8\alpha\gamma\kappa k\Phi'+r e^{\gamma\Phi}\right)\right. \nonumber \\
&&\left. +r e^{\gamma\Phi}\left\{4B\alpha\gamma\kappa^2(k-5B)\Phi'^2-\Lambda e^{\lambda\Phi} \left(-8 B\alpha\gamma\kappa +8\alpha\gamma\kappa k +\lambda r^2 e^{\gamma\Phi}\right) \right. \right. \nonumber \\
&&\left. \left. - 2B \kappa r\Phi'\left(e^{\gamma\Phi} +4\alpha\gamma  \lambda \Lambda e^{\lambda\Phi}\right)\right\}\right. \nonumber \\
&&\left. +B \kappa\delta' \left\{8\alpha\gamma (B-k) e^{\gamma\Phi}-8 B\alpha\gamma\kappa \Phi'^2 \left(8\alpha\gamma^2 (B-k)-r^2 e^{\gamma\Phi}\right)+r^3 e^{2\gamma\Phi} \Phi'\right\}\right]=0\,.  \label{DilatonEq}
\end{eqnarray}
In this paper, we choose Eqs.\ (\ref{gtt}), (\ref{DeltaEq}) and (\ref{DilatonEq}) as the equations of motion for our numerical computations.

We impose the boundary condition at the event horizon to guarantee the existence of a black hole. The event horizon is the hypersurface for $g^{rr}(r_h)=0$ or $g_{rr}(r_h)=\infty$. At the horizon, the metric function $B(r_h)=0$, the lapse function $\delta(r_h)$ is finite, and the scalar field $\Phi(r_h)$ is finite. From Eq.\ (\ref{gtt}) we obtain the value of $B'(r)$ at the horizon as follows:
\begin{equation}
B'(r_h)=\frac{e^{\gamma\Phi _h} (\Lambda r^2_h e^{\lambda\Phi _h}+k)}{r_h e^{\gamma\Phi _h}-4\alpha\gamma\kappa k\Phi'_h}\,. \label{brh}
\end{equation}
By plugging Eq.\ (\ref{brh}) into (\ref{DilatonEq}), we obtain a quadratic equation with respect to $\Phi'(r_h)$, which is related to $\Phi(r_h)$ as follows:
\begin{equation}
\Phi'(r_h) = \frac{-D \pm \sqrt{D^2 -4AC}}{2A} \,, \label{phi_rh_k}
\end{equation}
where
\begin{eqnarray}
A&=& 4\alpha\gamma\kappa [r^2_h e^{\gamma\Phi_h}(ke^{\gamma\Phi_h} - 4\alpha\gamma k \lambda\Lambda e^{\lambda\Phi_h}+ \Lambda r^2_h e^{\Phi_h(\gamma+\lambda)}) -32\alpha^2\gamma^2\kappa k^2 \Lambda e^{\lambda\Phi_h}  ] \,, \nonumber \\
D&=& r_h e^{\gamma\Phi_h} \left[ 96\alpha^2 \gamma^2 \kappa k \Lambda e^{\lambda\Phi_h} + r^2_h e^{\gamma\Phi_h} \left(-e^{\gamma\Phi_h} +32\alpha^2 \gamma^2 \kappa \Lambda^2 e^{\Phi_h(2\lambda-\gamma)} +8\alpha\gamma \lambda\Lambda e^{\lambda\Phi_h} \right. \right.   \nonumber \\
&&\left. \left.  - \Lambda r^2_h e^{\Phi_h(\gamma +\lambda)}/k \right) \right] \,,  \nonumber \\
C&=& \frac{e^{2\gamma\Phi_h}}{2\kappa} \left[24 \alpha\gamma\kappa  k +16\alpha\gamma\kappa \Lambda r^2_h e^{\lambda\Phi_h} -\frac{2\Lambda r^4_h}{k}\left(-4\alpha\gamma\kappa\Lambda e^{2\lambda\Phi_h} +\lambda e^{\Phi_h(\gamma +\lambda)}\right)\right]\,, \nonumber \label{phi_rh_coe}
\end{eqnarray}
and only one of two solutions provides a stable asymptotic black hole solution. We choose the $-$ sign in Eq.\ (\ref{phi_rh_k}). The term inside the square root of $\Phi'(r_h)$ should be positive; therefore, only some values in certain regions of $\Phi_h$ and $r_h$ can make $\Phi'(r_h)$ real. The values for $k=\pm 1$ have both allowed and forbidden regions which suggest that there are lower or upper bounds of the black hole size. If $\Lambda$ vanishes, it reduces to
\begin{equation}
\Phi'(r_h) = \frac{r_h e^{\gamma\Phi_h}}{8\alpha\gamma\kappa k}\left(1 \pm \sqrt{1 - 192 e^{-2\gamma\Phi_h}\alpha^2\gamma^2 \kappa k^2/r^4_h}\right) \,. \label{phi_rh_k_lamb}
\end{equation}
This one corresponds to $\Phi'(r_h)$ with $k$ shown in Ref.\ \cite{agll}. The near-horizon effects in $f(R)$ theories were studied in Ref.\ \cite{PerezBergliaffa:2011gj}.

We now impose appropriate boundary conditions for the behavior at the asymptotic region. We expect the metric function and the field to be
\begin{equation}
B(r) \simeq k - \frac{2GM}{r} + \frac{\Lambda e^{\lambda\Phi}}{3} r^2, ~~~~ \delta \simeq \delta_{\infty} + \frac{\delta_1}{r^{\beta}} ,  ~~~{\rm and}~~~   \Phi \simeq \Phi_{\infty} + \frac{Q}{r^{\sigma}}\,, \label{asympform}
\end{equation}
where $M$ is the black hole mass, $Q$ the scalar charge, and $\delta_{\infty}$ and $\Phi_{\infty}$ the asymptotic values of the lapse function and the scalar field, respectively, which will be used to rescale the lapse function, the scalar field, the radial coordinate, and the cosmological constant. The powers $\beta$ and $\sigma$ for $r$ are equal to $3$ as shown in Refs.\ \cite{got04, ohto04}. The mass of a hairy black hole is composed of the contributions both from the mass surrounded by the event horizon and from the existence of a scalar hair.

\subsection{Numerical solutions}

We now consider the symmetry and rescaling of some quantities. The scalar field is set to $\tilde {\Phi}=0$ in the asymptotic region, in which the value of $\Phi_{\infty}$ is determined by the values of $\alpha$, $\gamma$, $\lambda$, and $\Lambda$ \cite{got04, ohto04}. The lapse function is also set to $\tilde {\delta}=0$ in the asymptotic region, and then the geometry goes to an asymptotic AdS spacetime. Under these conditions, $\Phi$ is redefied by $\tilde {\Phi}=\Phi - \Phi_{\infty}$. To make the equations of motion invariant, some are rescaled as follows: $\tilde{\Lambda}=\Lambda e^{(\lambda-\gamma)\Phi_{\infty}}$, $\tilde{r}=re^{\gamma\Phi_{\infty}/2}$, $\tilde{t}=te^{\gamma\Phi_{\infty}/2}$, $\tilde{M}=Me^{\gamma\Phi_{\infty}/2}$.

We obtain the black hole solutions by solving Eqs.\ (\ref{gtt}), (\ref{DeltaEq}), and (\ref{DilatonEq}), generally. We take $\kappa =1$ for convenience. We first fix the couplings $\alpha$, $\gamma$, and $\lambda$. The values of $\Phi_{\infty}$ and $\Lambda$ are determined through $\Phi_{\infty}(\lambda-\gamma) = \ln\left(\frac{3\lambda}{4\alpha\gamma\Lambda}\right)$ as shown in Refs.\ \cite{got04, ohto04}. After fixing all couplings and $\Lambda$, we obtain the evolution of the tilde variables. From now on we use rescaled variables without a tilde. We first choose the initial value of $r_h$. The values of $\Phi_h$ are determined by requiring the term inside the square root to be positive. We divide numerical solutions into two types according to $k=\pm 1$.

\subsubsection{Spherically symmetric solutions with $k=1$}

\begin{figure}[t]
\begin{center}
\subfigure[Solution of the metric functions]
{\includegraphics[width =2.23 in]{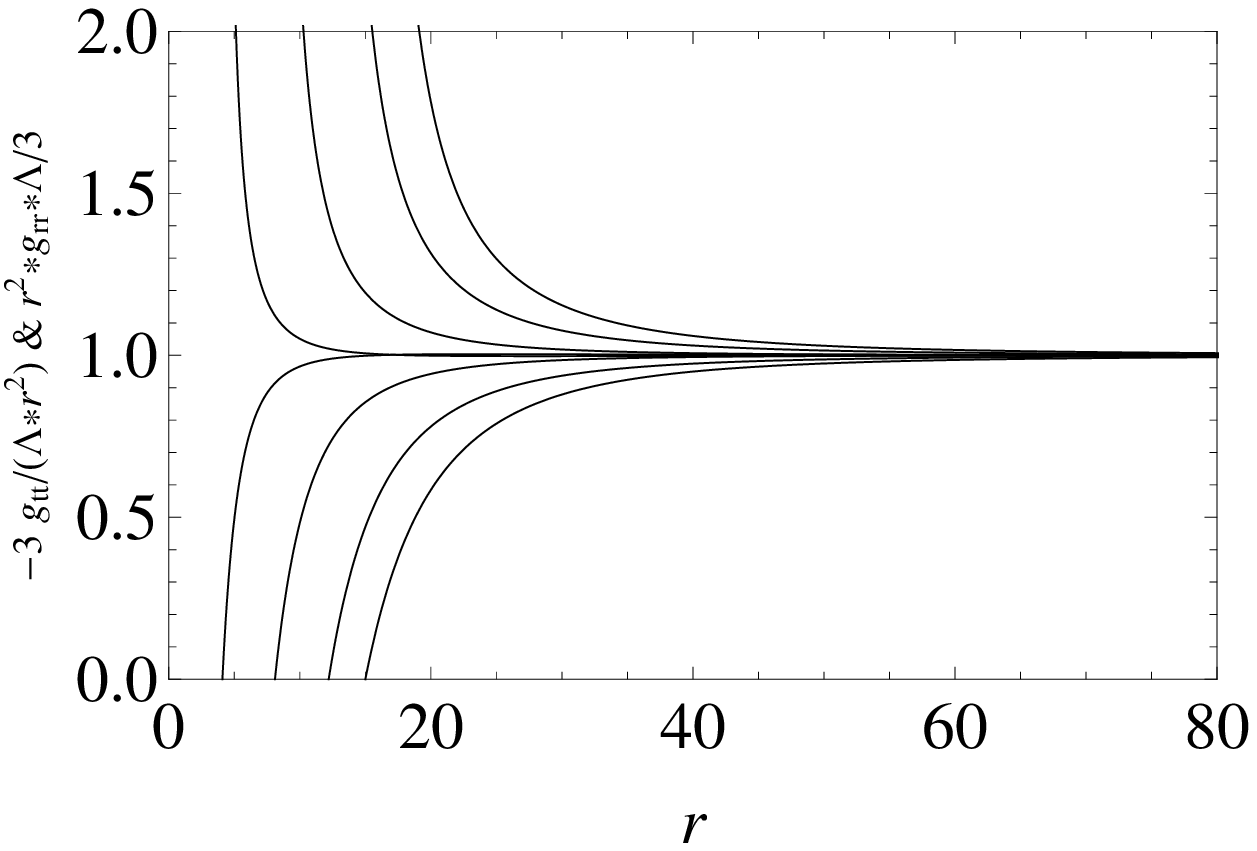}}
\subfigure[Solution of $\Phi(r)$]
{\includegraphics[width =2.4 in]{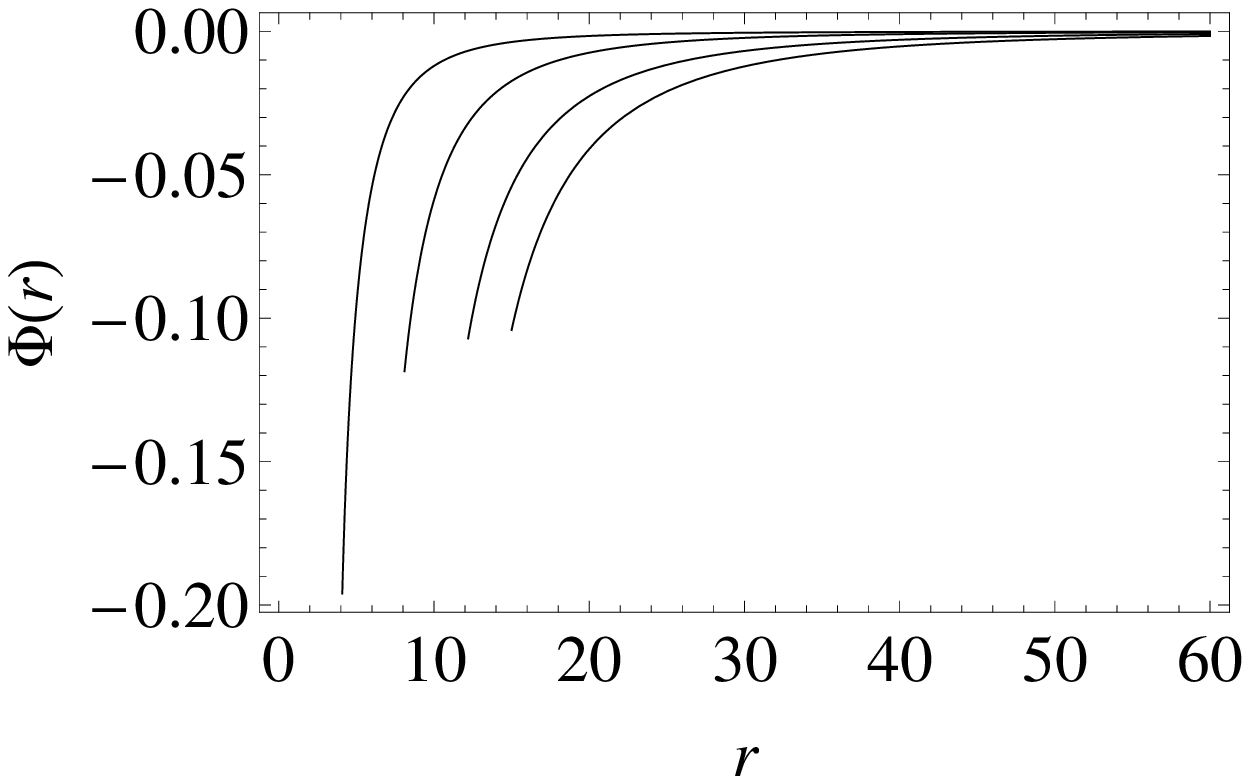}}\\
\subfigure[Solution of $\delta(r)$]
{\includegraphics[width =2.4 in]{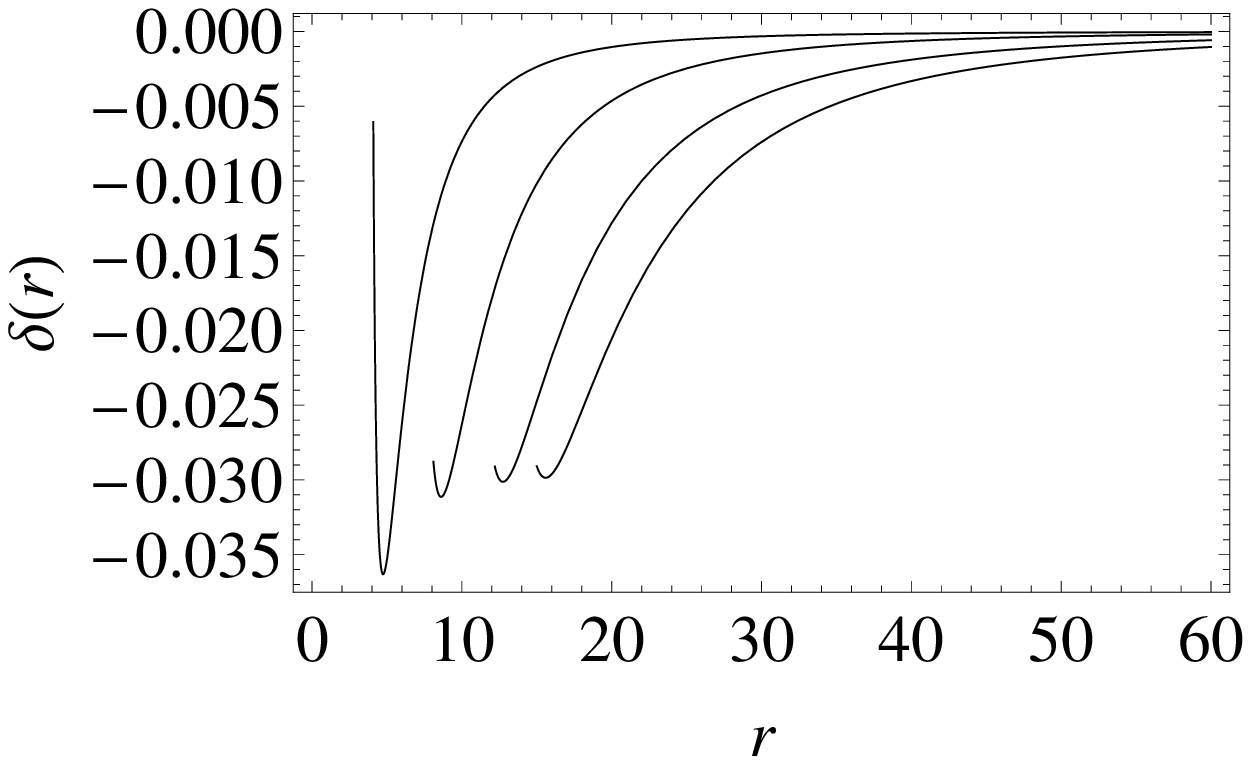}}
\subfigure[Mass function $m(r)$]
{\includegraphics[width =2.4 in]{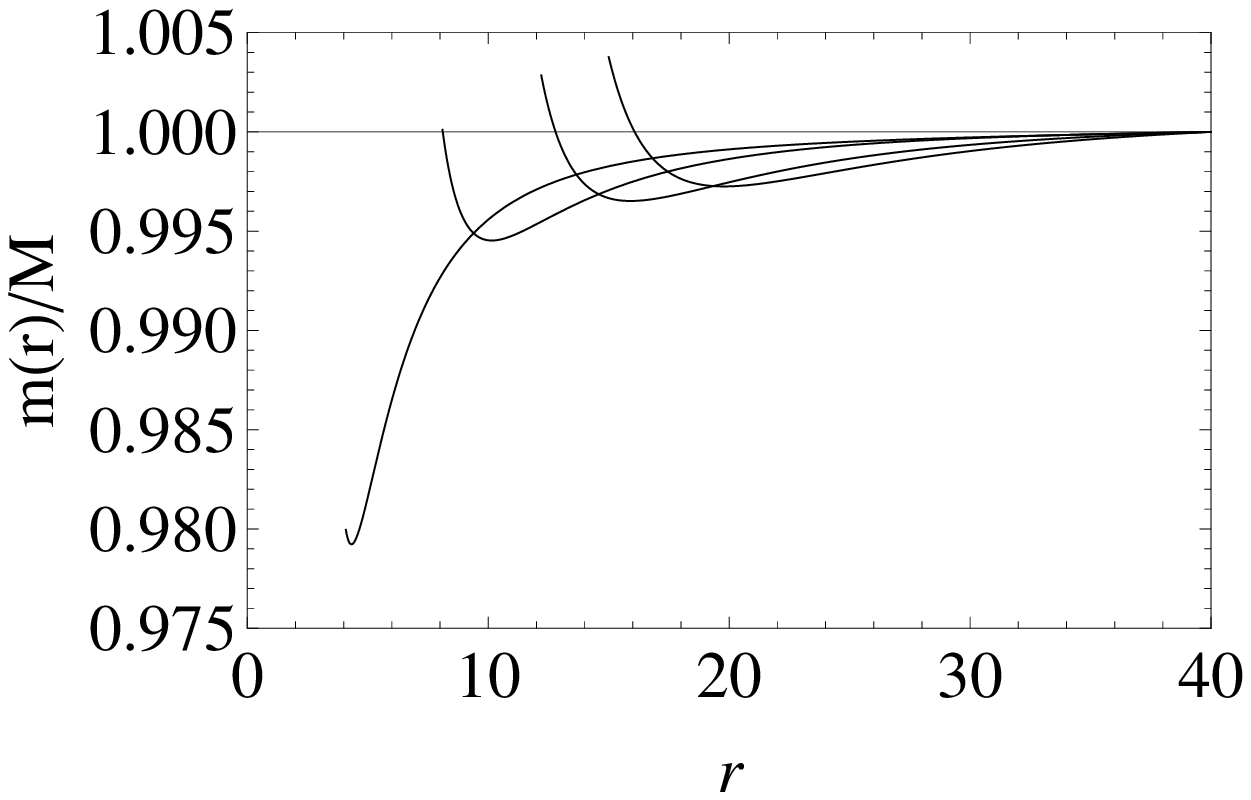}}
\end{center}
\caption{\footnotesize{(color online). Numerical solutions for $k=1$ with $\alpha=1$, $\gamma=1/2$, $\lambda=1/2$, $\Lambda=1/2$, and $\kappa=1$. We choose four solutions with the initial $r_h$ as $4.09$, $8.1$, $12.2$, and $15$, respectively.  }}
\label{Numerical-solutions-k+1}
\end{figure}

Figure \ref{Numerical-solutions-k+1} shows the numerical solutions for $k=1$. We take the parameter values as $\alpha=1$, $\gamma=1/2$, $\lambda=1/3$, and $\Lambda=1/2$ for Fig.\ \ref{Numerical-solutions-k+1}. We choose four solutions with the initial $r_h$ as $4.09$, $8.1$, $12.2$, and $15$, respectively. The solutions of the metric function are shown in Fig.\ \ref{Numerical-solutions-k+1}(a). The metric function $\frac{\Lambda}{3} r^2 g_{rr}$ diverges at the horizon, while $-g_{tt}/(\frac{\Lambda}{3}r^2)$ vanishes, whereas both functions approach the value $1$ asymptotically. The profiles of the scalar field are shown in Fig.\ \ref{Numerical-solutions-k+1}(b). The profiles start at the black hole horizon with negative values and approach zero monotonically. The end point of each profile indicates the value of the scalar field at the horizon as $\Phi_h$. The profiles of the lapse function are shown in Fig.\ \ref{Numerical-solutions-k+1}(c). The profiles start at the black hole horizon with negative values, rapidly decrease near the horizon, reach the minimum values, and then grow to zero at asymptotic infinity. Figure \ref{Numerical-solutions-k+1}(d) shows the mass function with different values of $r_h$. The mass function is numerically determined by integrating the equations of motion using the metric function, $-g_{tt}(r)=1-\frac{2Gm(r)}{r}+\frac{\Lambda e^{\lambda\Phi(r)}}{3}r^2$, according to Ref.\ \cite{ohto04}. The mass function decreases to the minimum value and increases up to some constant as the distance from the horizon increases in such a way that the ratio between the mass function and the asymptotic mass asymptotically approaches to $1$. The decreasing behavior for the mass ratios near the horizon $r_h$ shows that the contribution by the dilaton hair is negative, while the increasing behavior over the some $r$ shows that the contribution becomes positive.

\subsubsection{Hyperbolic solutions with $k=-1$}

\begin{figure}[t]
\begin{center}
\subfigure[Solution of the metric functions]
{\includegraphics[width =2.23 in]{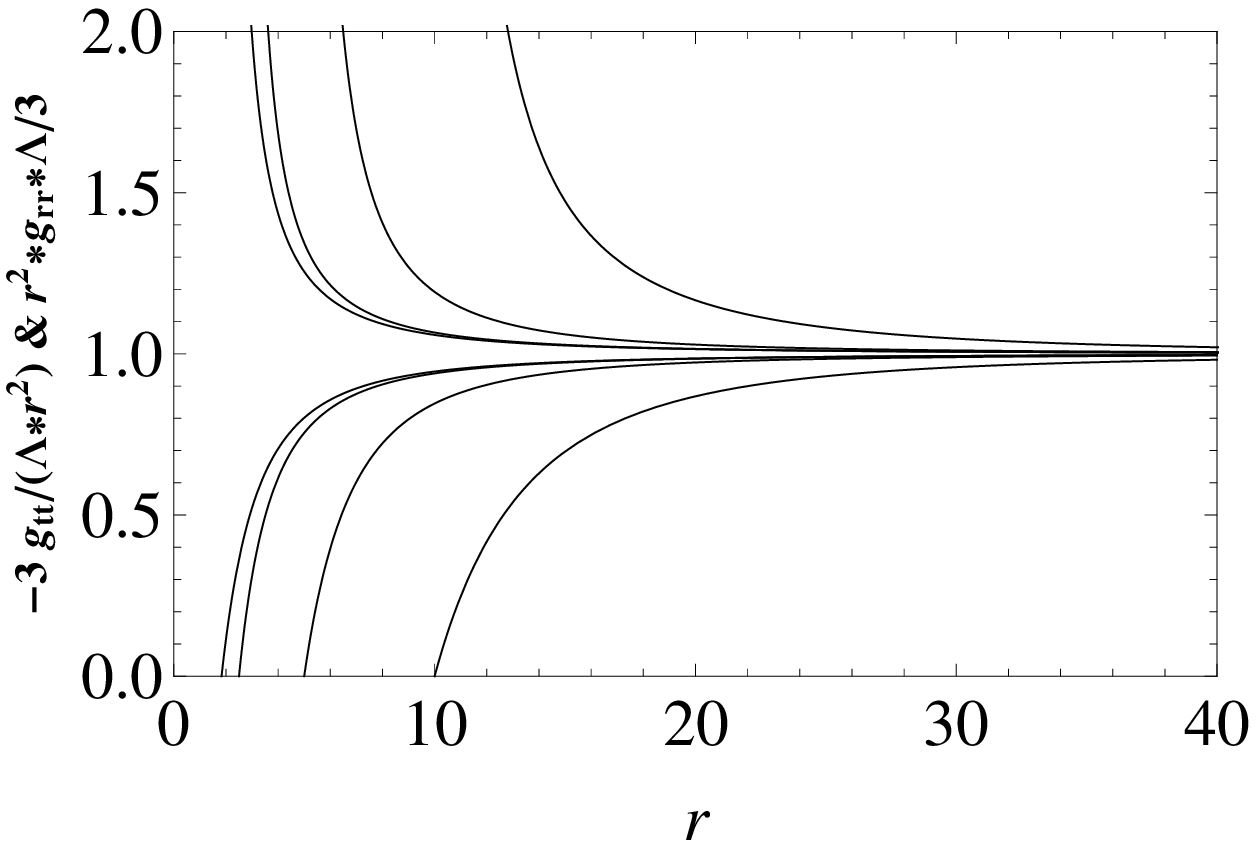}}
\subfigure[Solution of $\Phi(r)$]
{\includegraphics[width =2.4 in]{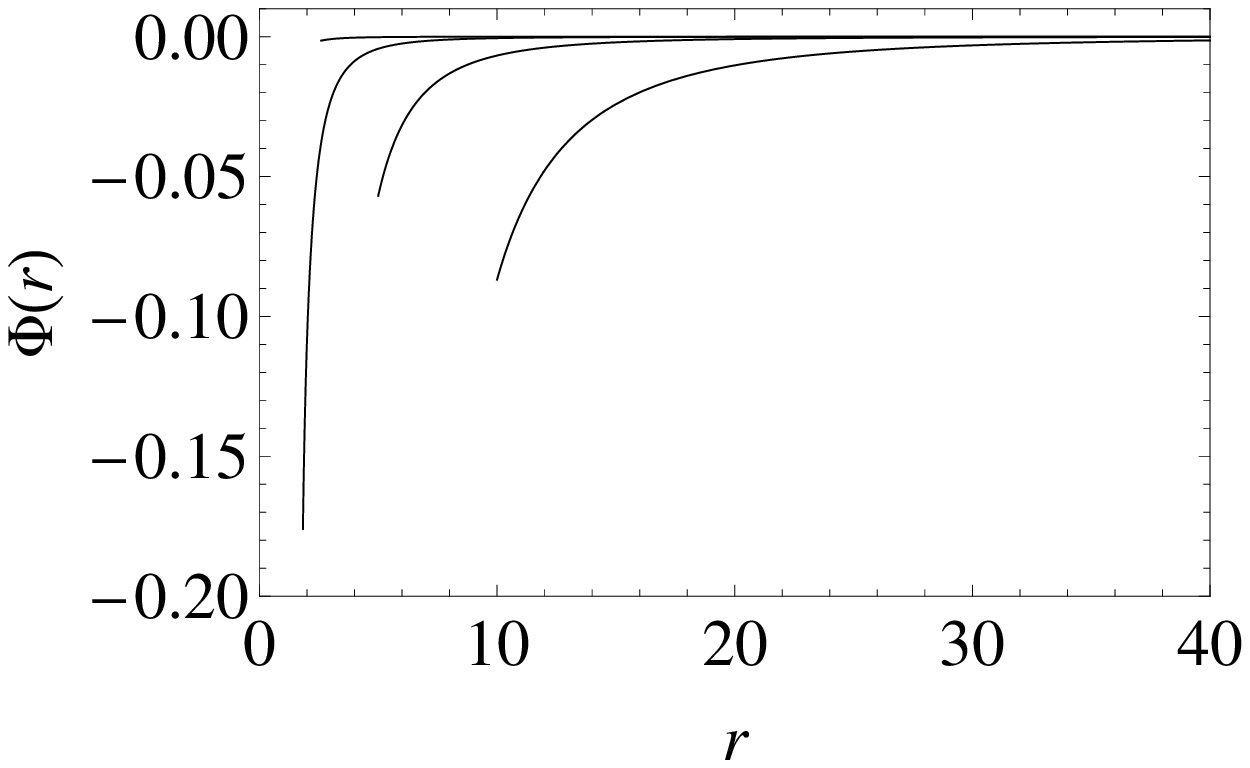}}\\
\subfigure[Solution of $\delta(r)$]
{\includegraphics[width =2.4 in]{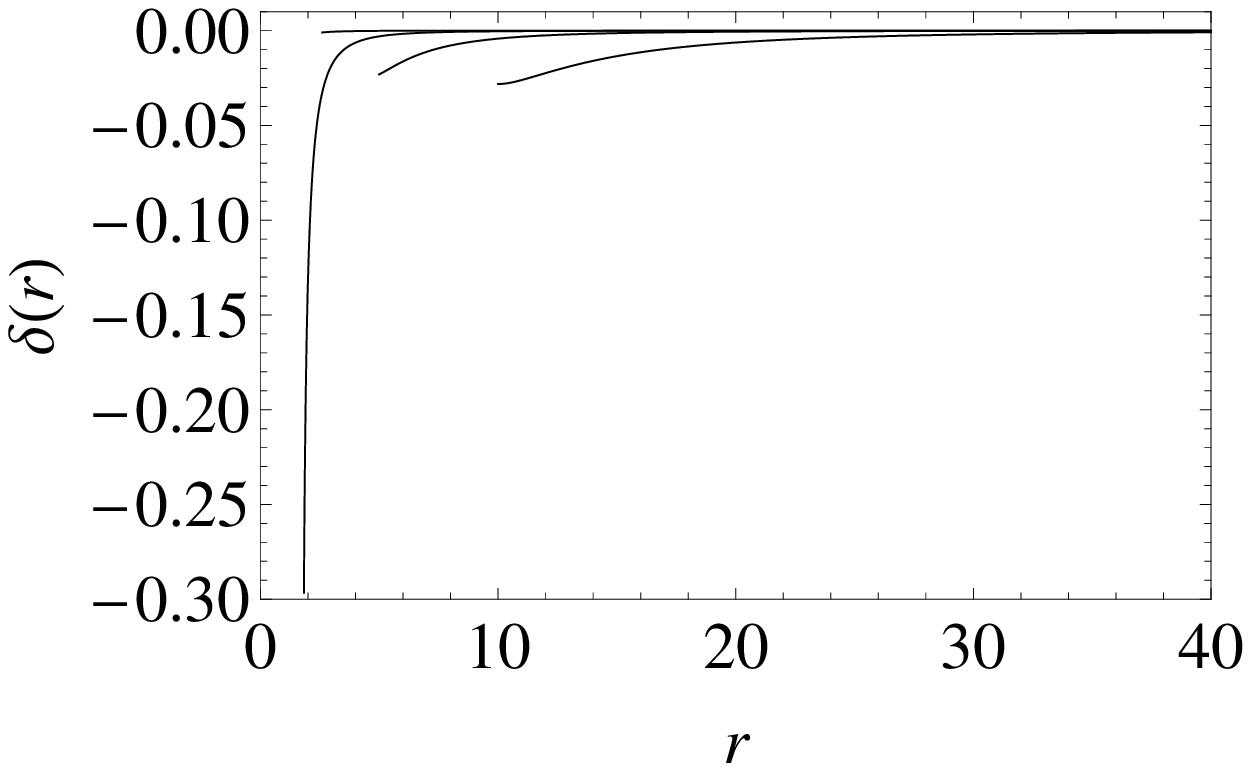}}
\subfigure[Mass function $m(r)$]
{\includegraphics[width =2.4 in]{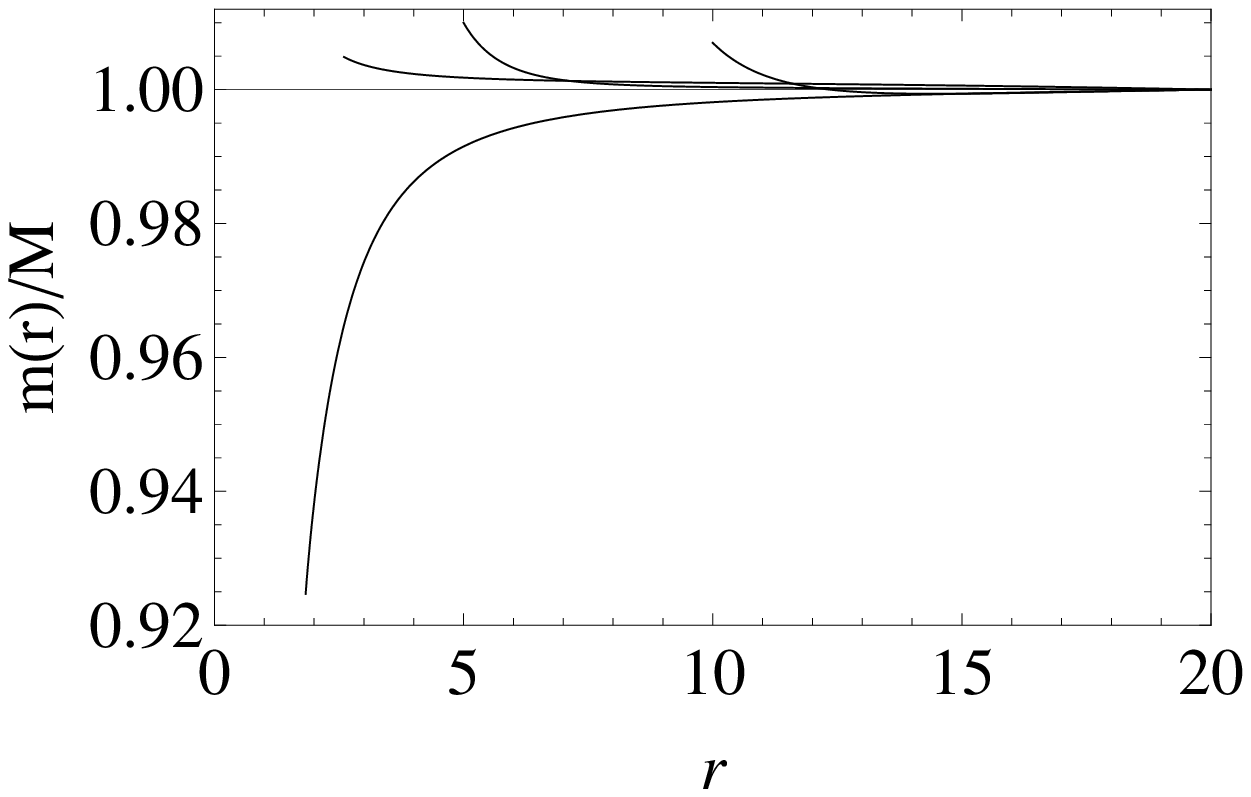}}
\end{center}
\caption{\footnotesize{(color online). Numerical solutions for $k=-1$  with $\alpha=1$, $\gamma=1/2$, $\lambda=1/2$, $\Lambda=1/2$, and $\kappa=1$. We choose four solutions with the initial $r_h$ as $1.83$, $2.60$, $5.00$, and $10.00$, respectively. }}
\label{Numerical-solutions-k-1}
\end{figure}

Figure \ref{Numerical-solutions-k-1} shows the numerical solutions for $k=-1$. We take the same parameter values used for Fig.\ \ref{Numerical-solutions-k+1}. We choose four solutions with the initial $r_h$ as $1.83$, $2.60$, $5.00$, and $10.00$, respectively. The solutions of the metric function are shown in Fig.\ \ref{Numerical-solutions-k-1}(a). This figure shows that the metric functions $\frac{\Lambda}{3} r^2 g_{rr}$ and $-g_{tt}/(\frac{\Lambda}{3}r^2)$ approach the value $1$ asymptotically. The profiles of the scalar field are shown in Fig.\ \ref{Numerical-solutions-k-1}(b). The lines start at the horizon and approach zero monotonically. The end points in each line make a convex curve. The first line with $r_h=1.83$ belongs to the left-hand side of the convex curve. The second line with $r_h=2.46$ belongs to about the maximum value of the convex curve. The third and fourth lines belong to the right-hand side of the convex curve. The profiles of the lapse function are shown in Fig.\ \ref{Numerical-solutions-k-1}(c). The end points in the lines also show similar behavior. The maximum values for $\Phi_h$ and $\delta_h$ are zero at $r^*_h=2.45$. Figure \ref{Numerical-solutions-k-1}(d) shows the mass function with different values of $r_h$, in which the mass function is numerically determined by using $-g_{tt}(r)=-1-\frac{2Gm(r)}{r}+\frac{\Lambda e^{\lambda\Phi(r)}}{3}r^2$. If the black hole is small the contribution from the hair is positive, while if the black hole becomes larger the contribution from the hair is negative to start with and is altered to be positive. The mass ratio is asymptotically approached to $1$. If the radius is at $r^*_h=2.45$ the mass ratio is exactly $1$ because of the null contribution by the hair, $\Phi(r^*_h)=0$.

\section{PHASE TRANSITION \label{sec3}}

We investigate black hole thermodynamic properties and the HP phase transition in (D)EGB theories. The study of black hole thermodynamics becomes an interesting subject after discovering the natural temperature and the intrinsic entropy of a black hole, in which the temperature is proportional to black hole's surface gravity and the entropy to one quarter of the area of its event horizon in Planck units. We explore how black hole thermodynamic properties and the phase transition can be affected by the higher-order curvature terms. The Euclidean path integral approach is one of the important tools to explore in this arena.

\subsection{Thermodynamic properties of AdS black holes}

We employ the partition function defined by a functional integral over all metrics and matter fields on the background \cite{HawPa}
\begin{equation}
Z=\int {\cal D}[g]{\cal D}[\Phi] \exp (-I_E[g, \Phi])  \,, \label{partition-function}
\end{equation}
where $I_E[g, \Phi]$ is the Euclidean action of gravitational fields and matter fields. In the semiclassical approximation, the dominant contribution to the path integral comes from classical solutions to the equations of motion. It can be evaluated by $\ln Z=-I_E$. In the canonical ensemble, the Helmholz free energy of black holes takes the form $F = M -T_H S$, in which the common redshift factor on both sides is eliminated and $T_H$ is the temperature corresponding to the inverse periodicity $T_H=\beta^{-1}_H$. One can use the relation $F =\frac{I_E}{\beta_H}$. The entropy is also derived from the relation $S=M\beta_H - I_E$. The temperature $T$ measured at spatial infinity is equal to the product of $T_H$ and the redshift factor, and the thermodynamic internal energy $E$ to the product of $M$ and the redshift factor. Thus, $E/T=M/T_H$. The conversed mass $M$ reduces to the mass of a black hole at spatial infinity. The specific heat determines the local stability of the thermodynamic system, which can be defined by $C=\frac{\partial M}{\partial T_H}$ \cite{bcma}.

The entropy formula should be modified to have the contribution from the GB term. From the first law of black hole thermodynamics, the entropy can have a constant term after the integration. The quantity can be nonzero and be determined in EGB theory. When the analytic form of the solution is known, we can obtain straightforwardly all thermodynamic quantities analytically after evaluating the Euclidean action. In EGB theory, we evaluate all thermodynamic quantities analytically. In DEGB theory, we make use of numerical computation.

\begin{figure}[t]
\begin{center}
\subfigure[Temperature for $k=1$]
{\includegraphics[width =2.5 in]{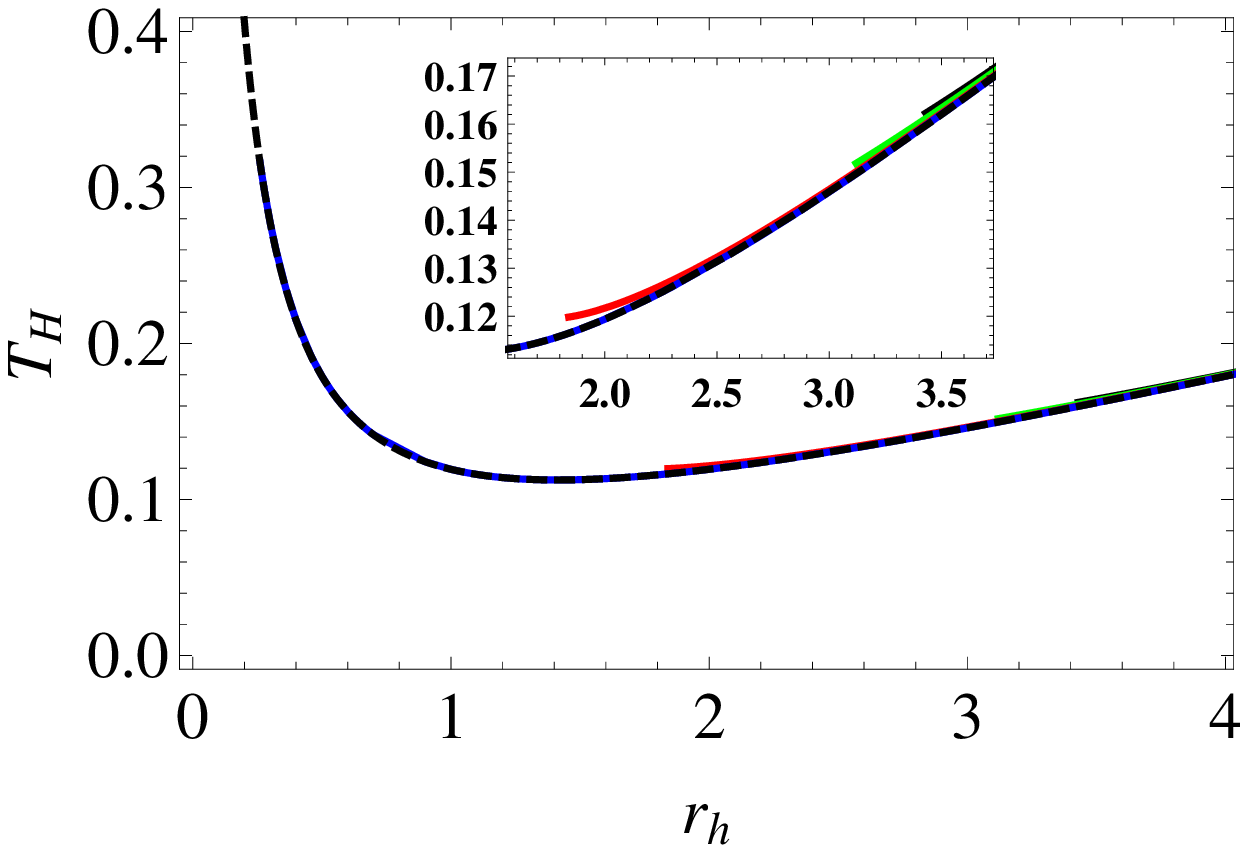}}
\subfigure[Temperature for $k=-1$]
{\includegraphics[width =2.5 in]{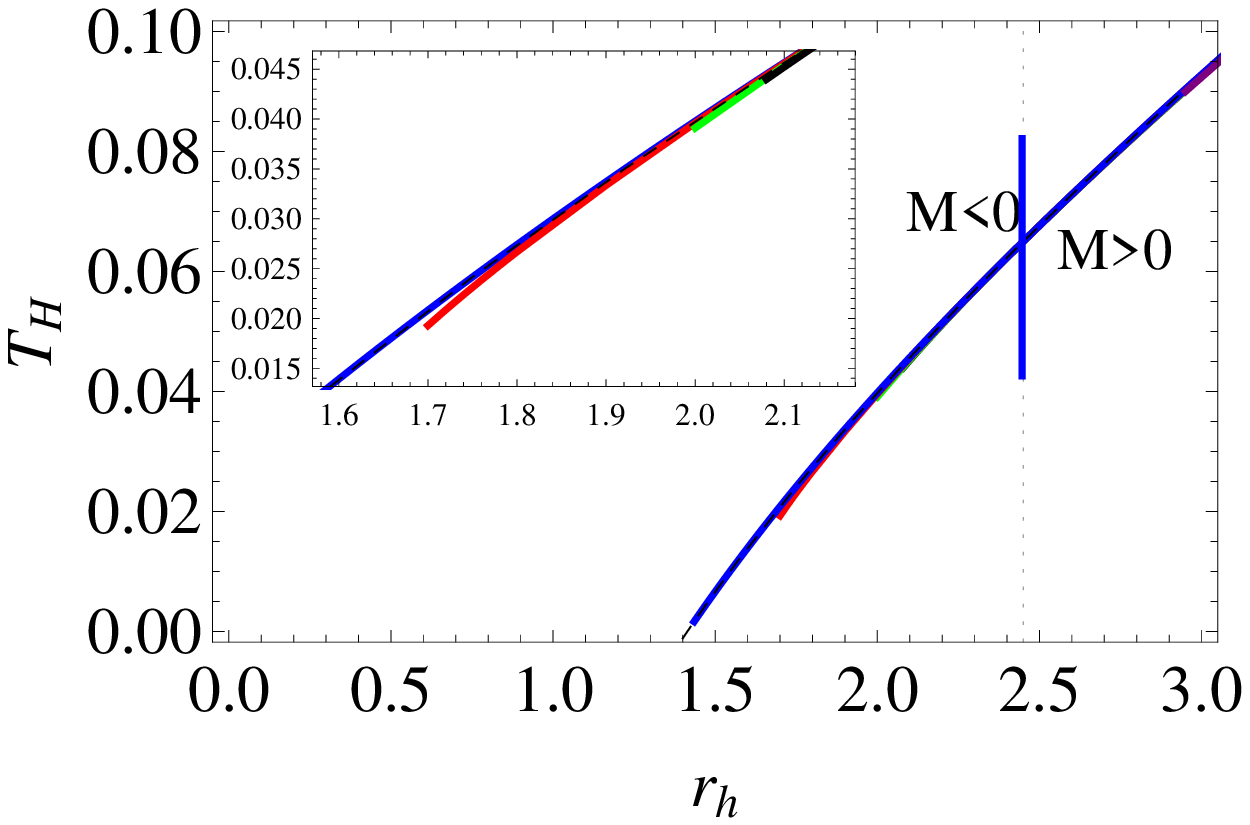}}
\end{center}
\caption{\footnotesize{(color online). The Hawking temperature as a function of the horizon radius $r_h$ with $\gamma=1/2$, $\Lambda=1/2$, and $\kappa=1$. $\alpha=0$ for the dashed line both in Einstein theory and EGB theory. $\alpha=0.005$ for the blue line, $\alpha=0.400$ for the red line, $\alpha=0.800$ for the green line, and $\alpha=1.000$ for the black line in DEGB theory.}}
\label{Hawking-temp}
\end{figure}

We begin by computing the temperature of a black hole. To avoid the conical singularity, we should impose the periodicity in the Euclidean black hole metric
\begin{equation}
\beta_H= \frac{4\pi}{B'(r_h)e^{-\delta(r_h)}} \,. \label{h-temperature}
\end{equation}
If $\delta(r)$, $\lambda$, and $\gamma$ vanish as both in Einstein and in EGB theory, the temperature takes the form $T_H =\frac{1}{2\pi} (\frac{GM}{r^2_h}+\frac{\Lambda}{3}r_h)$. We analyze the temperatures depending on the sign of $k$ in both theories. For $k=1$, we consider only the positive mass, because there is a naked singularity for the negative mass black hole. Then the temperature is always positive with the minimum value of $\sqrt{\Lambda}/2\pi$ at $r_h=1/\sqrt{\Lambda}$. For $k=-1$, there are two types of masses without the naked singularity. One is a positive mass and the other is a negative mass. For the positive mass black hole, the temperature is also positive. However, there is the limitation for the horizon radius, $r_h > \sqrt{3/\Lambda}$. For the negative mass black hole, there exist two horizons. When $1/\Lambda=9G^2M^2$, two horizons coincide at $r_{\rm crit}=1/\sqrt{\Lambda}$ and the temperature vanishes. This one corresponds to the extremal black hole with the zero temperature. However, there is the limitation for the horizon radius, $r_h < \sqrt{3/\Lambda}$. To treat both cases simultaneously, we take the form $ T_H =\frac{1}{4\pi r_h}\left(k+\Lambda r^2_h\right)$ by using $M(r_h)=\frac{r_h}{2G}\left(k+\frac{\Lambda}{3} r^2_h \right)$. Then $\beta_H= \frac{4\pi r_h}{k+\Lambda r^2_h}$. The mass for the extremal black hole is given by $M_{\rm crit}=-\frac{1}{3G\sqrt{\Lambda}}$ \cite{robm, birm, sugo, agpa}.

Figure \ref{Hawking-temp} shows the temperature in DEGB theory. We take $\gamma=1/2$, $\Lambda=1/2$, and $\kappa=1$. The end point in each line indicates the minimum mass of a black hole with given parameter values. The dashed line corresponds to the case both in Einstein and in EGB theory. The blue line corresponds to $\alpha=0.005$ and $\lambda=1/600$, the red line to $\alpha=0.400$ and $\lambda=2/15$, the green line to $\alpha=0.800$ and $\lambda=4/15$, and the black line to $\alpha=1.000$ and $\lambda=1/3$. Figure \ref{Hawking-temp}(a) shows the temperature as a function of the horizon radius $r_h$ for $k=1$. The minimum temperature, the minimum radius, and the low bound of the black hole mass increases as the value of $\alpha$ increases. The end points are $r_h=0.26$ for the blue line, $r_h=1.84$ for the red line, $r_h=3.12$ for the green line, and $r_h=3.43$ for the black line, respectively. Figure \ref{Hawking-temp}(b) shows the temperature as a function of the horizon radius $r_h$ for $k=-1$. The end points are $r_h=1.44$ for the blue line, $r_h=1.65$ for the red line, $r_h=2.00$ for the green line, and $r_h=2.14$ for the black line, respectively. The upper left inset in the figure shows the magnification of a small region to distinguish the overlapped lines. The regions are divided into two parts. The first part belongs to the negative mass black hole and the second part to the positive mass black hole. We cut off the part with the negative temperature. The zero temperature corresponds to the extremal case.

\begin{figure}[t]
\begin{center}
\subfigure[Specific heat for $k=1$]
{\includegraphics[width =2.5 in]{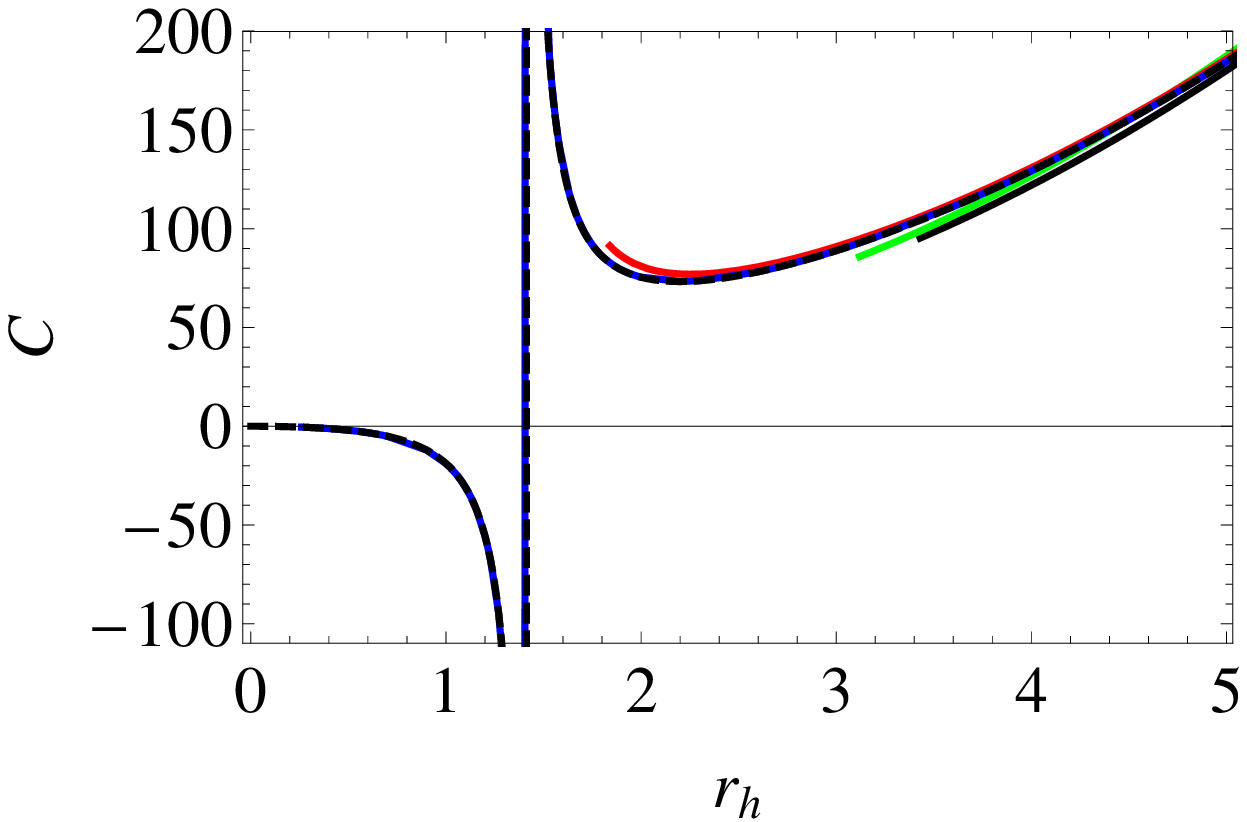}}
\subfigure[Specific heat for $k=-1$]
{\includegraphics[width =2.35 in]{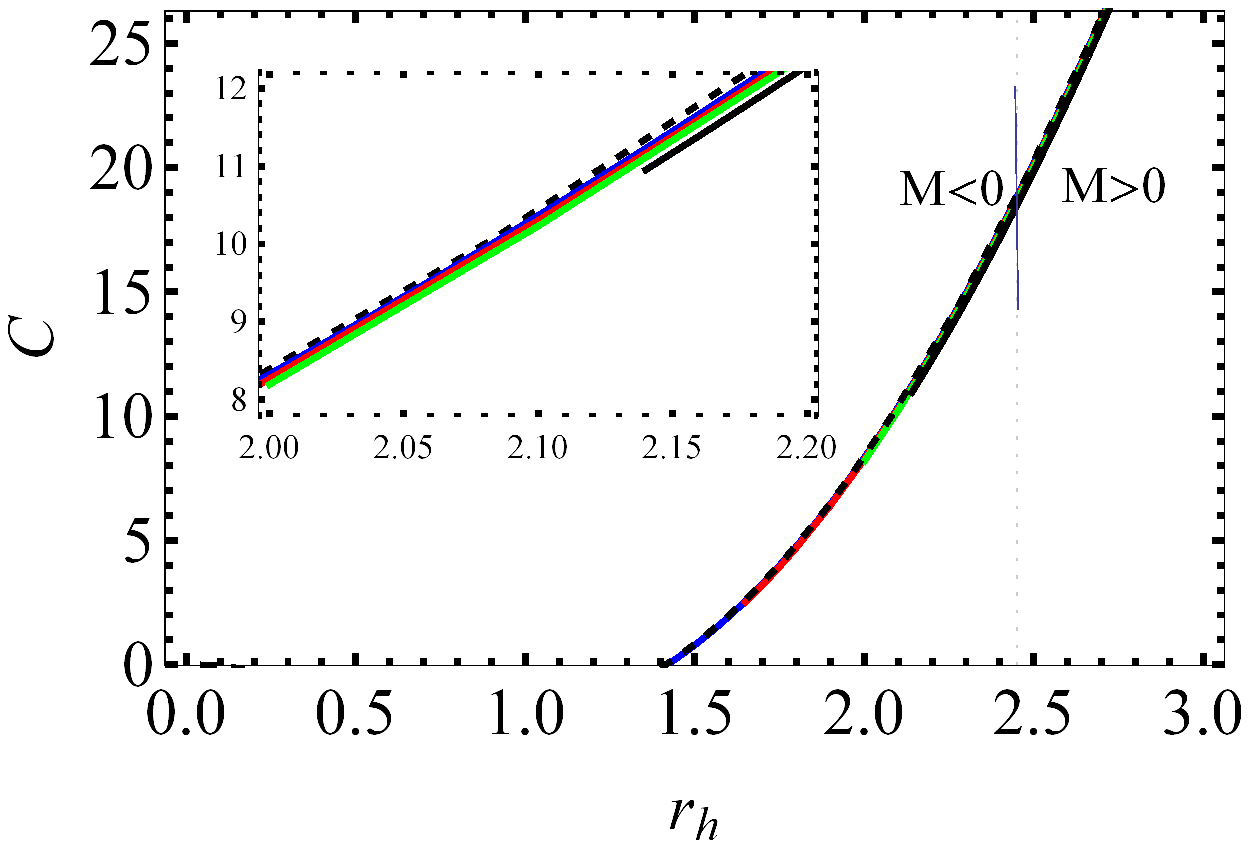}}
\end{center}
\caption{\footnotesize{(color online). The specific heat as a function of the horizon radius $r_h$ with $\gamma=1/2$, $\Lambda=1/2$, and $\kappa=1$. $\alpha=0$ for the dashed line both in Einstein theory and EGB theory. $\alpha=0.005$ for the blue line, $\alpha=0.400$ for the red line, $\alpha=0.800$ for the green line, and $\alpha=1.000$ for the black line in DEGB theory.}}
\label{Specific-heat}
\end{figure}

The specific heat can be obtained by using the relation $C=\frac{\partial M}{\partial T_H}$, in which $M$ is the mass of a black hole. If $\delta(r)$, $\lambda$, and $\gamma$ vanishes, the specific heat takes the form $C=\frac{2\pi}{G}\frac{r^2_h(k+\Lambda r^2_h)}{(\Lambda r^2_h-k)}$. The specific heat determines the local stability of a thermodynamic system. The behavior of the change from the negative to positive specific heat describes the existence of local stability depending on the size of a black hole as well as justifies employing the canonical ensemble description.

Figure \ref{Specific-heat} shows the specific heat in DEGB theory. We take the same parameters used for Fig.\ \ref{Hawking-temp}. The dotted line corresponds to the reference line, which is obtained for the specific heat both in Einstein and in EGB theory. Figure \ref{Specific-heat}(a) shows the specific heat as a function of the horizon radius $r_h$ for $k=1$. Figure \ref{Specific-heat}(b) shows the specific heat as a function of the horizon radius $r_h$ for $k=-1$. The upper left inset in the figure shows the magnification of a small region to distinguish the overlapped lines. We cut off the data corresponding to the negative temperature.

We now consider the Euclidean action. We can evaluate analytically the Euclidean action only both in Einstein theory and EGB theory, because we do not know the analytic solution in DEGB theory. The total Euclidean action can be divided into the following:
\begin{equation}
I_E({\rm total}) = I_E({\rm bulk}) + I_E({\rm YGH}) + I_E({\rm GBb}) + I_E({\rm  counterterm}) \,.
\end{equation}

We first review a black hole in AdS space in Einstein gravity. For the action of a black hole in AdS space, the first two terms give $I^{\rm BE}_E({\rm bulk}) + I^{\rm BE}_E({\rm YGH}) = x\beta_H(- \frac{\Lambda}{6G} r^3_{co} - \frac{\Lambda}{12G} r^3_h - \frac{k r_{co}}{2G} + \frac{3 M}{4} )$, in which the cutoff $r_{co}$ is going to $\infty$, $x$ is equal to $2$ for $k=1$, and $(\cosh\psi_{co} -1)$ for $k=-1$. $(\cosh\psi_{co} -1)$ denotes the integrated quantity for $\psi$ from $0$ to $\infty$. In this paper we take $x=2$ as well as the case for $k=-1$. We used $\sqrt{g_{\tau\tau}} \simeq \sqrt{\frac{\Lambda}{3}} r_{co} \left( 1 + \frac{3k}{2\Lambda r^2_{co}} - \frac{3GM}{\Lambda r^3_{co}} \right)$ to obtain the Euclidean action. The superscript ``BE" means the black hole solution in Einstein gravity. According to Ref.\ \cite{bakr}, the last term gives $I^{\rm BE}_E({\rm  counterterm})=x \beta_H (-\frac{M}{2} + \frac{k r_{co}}{2G} + \frac{\Lambda r^3_{co}}{6G})$. Then the total action for a black hole in AdS space gives the finite terms $I^{\rm BE}_E({\rm total}) =x\beta_H(\frac{M}{4} - \frac{\Lambda}{12G} r^3_h)$. We make use of the thermodynamic relation between the entropy and the Euclidean action $S^{\rm BE}=\frac{x M\beta_H}{2} - I^{\rm BE}_E({\rm total})$. Then the entropy of a black hole in AdS space is given by $S^{\rm BE}_E=\frac{A}{4G}$, in which $A$ is equal to $4\pi r^2_h$ for $k=1$ and $2\pi(\cosh\psi_{co} -1)r^2_h$ for $k=-1$.

We consider the black hole in AdS space in EGB theory. The action of the GB term is evaluated to be
\begin{equation}
\int_{\mathcal M} \sqrt{g_E} d^4 x_E \left[ -\alpha(R^2 - 4R_{\mu\nu}R^{\mu\nu} + R_{\mu\nu\rho\sigma}R^{\mu\nu\rho\sigma}) \right] =x \left( - 32k\pi^2 \alpha + \frac{\alpha\beta_H 16\pi \Lambda }{3}(GM- \frac{\Lambda}{3} r^3_{co}) \right) \,, \label{gbinte}
\end{equation}
where we plugged $R=-4\Lambda$, $R_{\mu\nu}R^{\mu\nu}=4\Lambda^2$, and $R_{\mu\nu\rho\sigma}R^{\mu\nu\rho\sigma}=\frac{48 G^2 M^2}{r^6}+\frac{8\Lambda^2}{3}$. We now consider the boundary term, $I_E({\rm GBb})$, shown in \cite{Myers:1987yn, Brihaye:2008xu}. This boundary term turns out to be zero for the black hole in the asymptotically flat spacetime. However, the boundary term has nonzero contributions to cancel out the terms with $\Lambda$ in Eq.\ (\ref{gbinte}) for the black hole in the asymptotically AdS spacetime. For our case it takes the form,
\begin{equation}
I_E({\rm GBb})= -x\alpha\beta_H 16\pi \left[\frac{2G^2M^2}{r^3_{co}} + \frac{\Lambda}{3}\left(GM- \frac{\Lambda}{3} r^3_{co}\right) \right] \,.
\end{equation}

\begin{figure}[t]
\begin{center}
\subfigure[Entropy for $k=1$]
{\includegraphics[width =2.5 in]{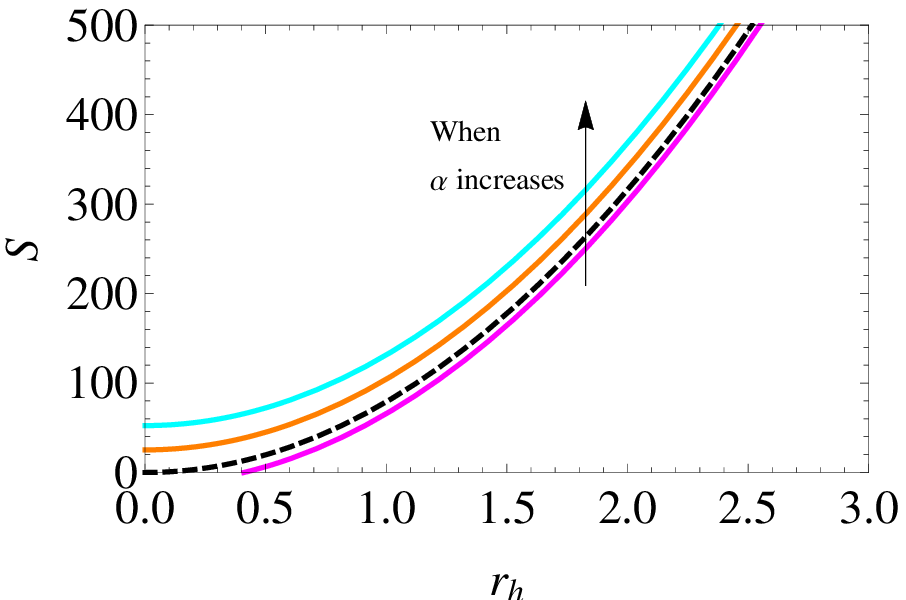}}
\subfigure[Entropy for $k=-1$]
{\includegraphics[width =2.5 in]{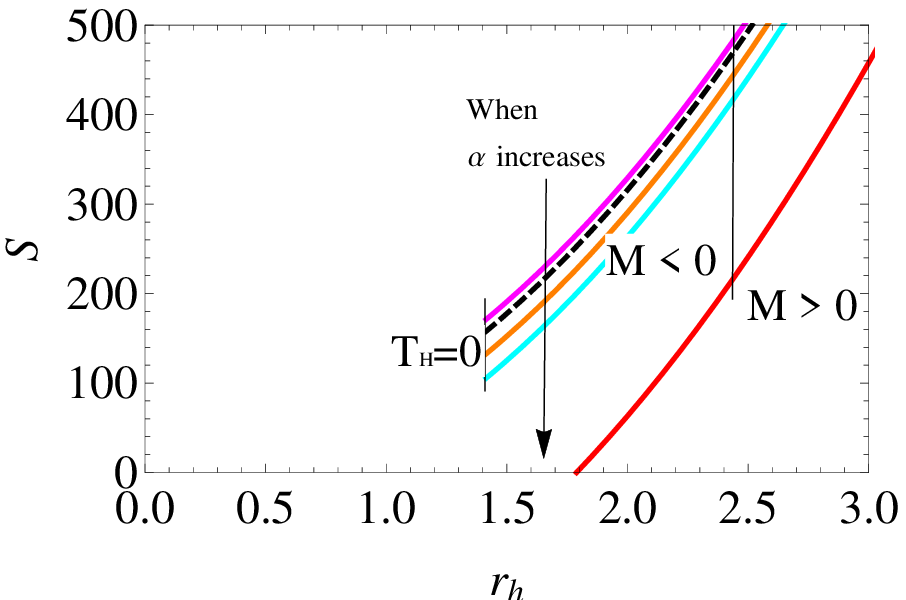}}
\end{center}
\caption{\footnotesize{(color online). The entropy as a function of the horizon radius $r_h$ in EGB theory with $\Lambda=1/2$, and $\kappa=1$. $\alpha=0$ for the dashed line in Einstein theory. $\alpha=0.040$ for the orange line, $\alpha=0.083$ for the cyan line, $\alpha=0.400$ for the red line, and $\alpha=-0.021$ for the magenta line in EGB theory.}}
\label{Entropy-EGB}
\end{figure}

The total action for a black hole in EGB theory is
\begin{equation}
I^{\rm BEGB}_E = x\left( \frac{\beta_H M}{4} - \frac{\beta_H\Lambda}{12G} r^3_h - 32k\pi^2 \alpha \right) \,,
\end{equation}
where the superscript ``BEGB" means the black hole solution in EGB theory. The third term in the right-hand side is a constant term depending on the sign and magnitude of $\alpha$, which provides the information on the topology of that spacetime manifold. The entropy is given by
\begin{equation}
S^{\rm EGB} =  \frac{A}{4G} \left(1+ \frac{8k\alpha\kappa}{r^2_h} \right) \,,
\end{equation}
where we can restrict the value of $\alpha$. From the positiveness of the entropy, the available values are given by $\alpha > - \frac{r^2_h}{8 \kappa}$ for $k=1$ and $\alpha < \frac{r^2_h}{8 \kappa}$ for $k=-1$. This entropy formula with a correction has the same form obtained by the different method as shown in Refs.\ \cite{jamy, sawa}.

Figure \ref{Entropy-EGB} shows the entropy in EGB theory. We take $\Lambda=1/2$ and $\kappa=1$. The black dashed line corresponds to $\alpha=0.000$ representing the entropy in Einstein theory. The orange line corresponds to $\alpha=0.040$, the cyan line to $\alpha=0.083$, the red line to $\alpha=0.400$, and the magenta line to $\alpha=-0.021$ with the positive entropy above $r_h > \sqrt{0.168}$. Figure \ref{Entropy-EGB}(a) shows the entropy as a function of the horizon radius $r_h$ for $k=1$. We cut off the part with the negative entropy. The entropy increases as the value of $\alpha$ increases. Figure \ref{Entropy-EGB}(b) shows the entropy as a function of the horizon radius $r_h$ for $k=-1$. The region is divided into two parts, the positive mass and the negative mass. The entropy decreases as the value of $\alpha$ increases.

We now consider the black hole in AdS space in DEGB theory. The entropy is obtained by numerical computation, because we cannot evaluate the Euclidean action and the counterterm analytically. We adopt the entropy formula for the black hole in DGB theory \cite{toma3}, which is given by
\begin{equation}
S^{DEGB}= \frac{A}{4G} \left(1+ \frac{8k\alpha\kappa}{r^2_h}e^{-\gamma\Phi_h} \right)\,.
\end{equation}
For the positiveness of the entropy, the value of $\alpha$ is restricted to $\alpha > - \frac{r^2_h}{8\kappa} e^{\gamma\Phi_h}$ for $k=1$ and $\alpha <  \frac{r^2_h}{8\kappa} e^{\gamma\Phi_h}$ for $k=-1$.

\begin{figure}[t]
\begin{center}
\subfigure[Entropy for $k=1$]
{\includegraphics[width =2.5 in]{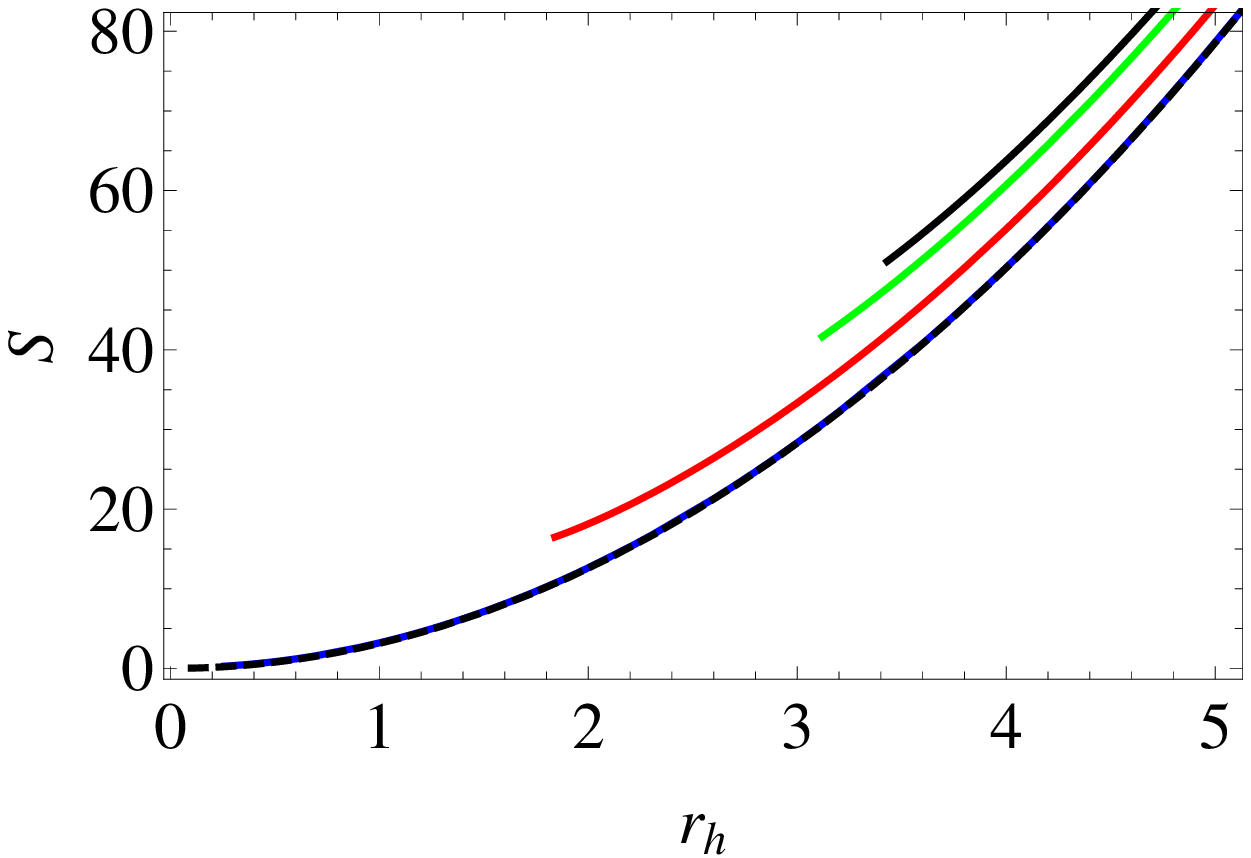}}
\subfigure[Entropy for $k=-1$]
{\includegraphics[width =2.5 in]{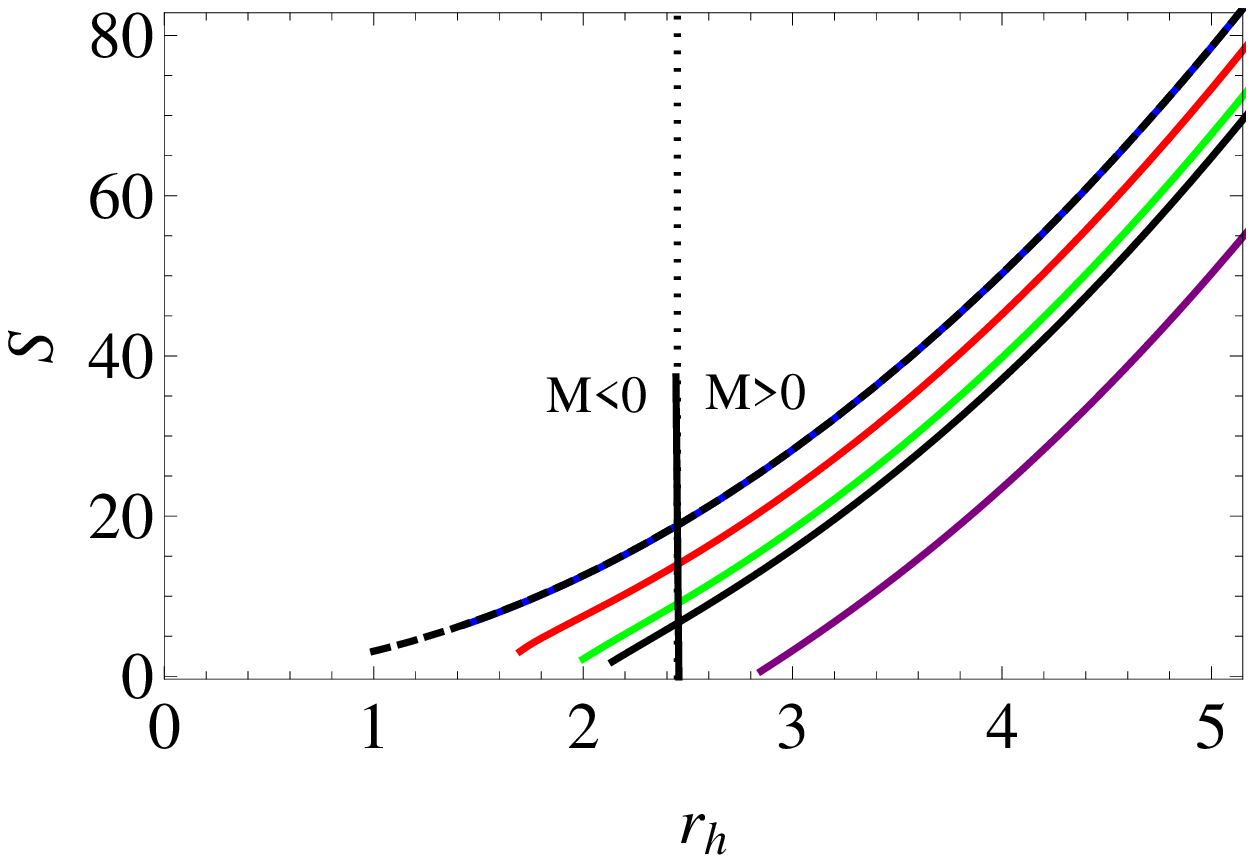}}
\end{center}
\caption{\footnotesize{(color online). The entropy as a function of the horizon radius $r_h$ with $\gamma=1/2$, $\Lambda=1/2$, and $\kappa=1$. $\alpha=0.000$ for the dashed line in Einstein theory. $\alpha=0.005$ for the blue line, $\alpha=0.400$ for the red line, $\alpha=0.800$ for the green line, $\alpha=1.000$ for the black line, and $\alpha=2.000$ for the purple line in DEGB theory.}}
\label{Entropy-DEGB}
\end{figure}

Figure \ref{Entropy-DEGB} shows the entropy in DEGB theory. We feed the numerical data into the formula. We take $\gamma=1/2$ and $\Lambda=1/2$. The dashed line corresponds to the case in Einstein theory. The blue line corresponds to the case with $\alpha=0.005$ and $\lambda=1/600$, the red line to the case with $\alpha=0.400$ and $\lambda=2/15$, the green line to the case with $\alpha=0.800$ and $\lambda=4/15$, the black line to the case with $\alpha=1.000$ and $\lambda=1/3$, and the purple line to the case with $\alpha=2$ and $\lambda=2/3$. Figure \ref{Entropy-DEGB}(a) shows the entropy as a function of the horizon radius $r_h$ for $k=1$. Figure \ref{Entropy-DEGB}(b) shows the entropy as a function of the horizon radius $r_h$ for $k=-1$. The vertical dotted line roughly divides the mass between the negative and positive mass.

\subsection{Hawking-Page phase transition}

We consider the Hawking-Page phase transition between a black hole in AdS space with $M\neq 0$ and the thermal AdS space with $M=0$ for $k=1$, and with $M=M_{\rm crit}$ for $k=-1$ \cite{vanzo, birm, cpark5} as a first-order phase transition \cite{HawPa}.

To obtain the free energy difference we use the thermodynamic relation. First, we consider the transition in Einstein gravity. The total Euclidean action is given by $I^{\rm BE}_E({\rm total}) =x(\frac{\beta_H M}{4} - \frac{\Lambda\beta_H}{12G} r^3_h)$. For the action of the thermal AdS space, the first two terms give
\begin{equation}
I^{\rm TE}_E({\rm bulk}) + I^{\rm TE}_E({\rm YGH}) = x \beta_o \left[- \frac{\Lambda }{6G} r^3_{co} - \frac{k r_{co}}{2G} +\frac{1-k}{2}\left(-\frac{\Lambda}{12G} r^3_{\rm crit} + \frac{3 M_{\rm crit}}{4}  \right)\right] \,,
\end{equation}
where the superscript ``TE" means the thermal AdS in Einstein gravity. We take the reference background geometry as the AdS black hole geometry with the critical mass for $k=-1$, in which the spacetime region is restricted to $r\geq r_{\rm crit}(=\sqrt{1/\Lambda})$. Because the AdS geometry is not associated with its natural temperature for $k=1$ and the geometry with the critical mass has the vanishing temperature for $k=-1$, one can introduce any temperature for the reference background geometry by suitably choosing the period of Euclidean time. The periodicity $\beta_o$ is satisfied with the condition, $\beta_o \sqrt{k+ \frac{\Lambda}{3}r^2-\frac{(1-k)GM_{\rm crit}}{r}}=\beta_H \sqrt{k-\frac{2GM}{r} + \frac{\Lambda}{3} r^2}$. And the total action with the counterterm is given by $ I^{\rm TE}_E({\rm total}) = x \frac{1-k}{2} \beta_o \left(-\frac{\Lambda}{12G} r^3_{\rm crit} + \frac{ M_{\rm crit}}{4}  \right) \equiv \frac{x(1-k)}{4}\beta_o M_{\rm crit} $. In other words, one introduce the counterterm to give the vanishing action for $k=1$ and the constant value for $k=-1$.

We mention about the reference background geometry in Einstein theory. For $k=1$, the action, the entropy, the mass, and the free energy are all equal to zero. For $k=-1$, the entropy is equal to zero, which can be easily calculated by $S=\beta_o M - I^{\rm TE}_E =0$. One can redefine the mass as $M'=M-M_{\rm crit}=0$ and the action as $I'^{\rm TE}_E = I^{\rm TE}_E - \frac{x}{2}\beta_o M_{\rm crit}=0$ \cite{birm}. In this paper we take the original action.

The difference of two actions given in Ref.\ \cite{HawPa} with $k$ is as follows:
\begin{equation}
I^{\rm E}_E = I^{\rm BE}_E({\rm total}) - I^{\rm TE}_E({\rm total})= x\left( \frac{\pi r^2_h(k-\frac{\Lambda}{3}r^2_h)}{2G(k+\Lambda r^2_h)} + \frac{\beta_o(1-k)}{12G\sqrt{\Lambda}}  \right)  \, .
\end{equation}
The free energy difference between a black hole in AdS and the thermal AdS is given by
\begin{equation}
F^E=  x\left( \frac{r_h(k-\frac{\Lambda}{3}r^2_h)}{8G} + \frac{(1-k)}{12G\sqrt{\Lambda}} \right) \,.
\end{equation}

Now we consider the transition in EGB theory. For the thermal AdS, the action of the GB term is
\begin{eqnarray}
&&\int_{\mathcal M} \sqrt{g_E} d^4 x_E \left[ -\alpha(R^2 - 4R_{\mu\nu}R^{\mu\nu} + R_{\mu\nu\rho\sigma}R^{\mu\nu\rho\sigma}) \right] \\ \nonumber
&& =  x \alpha\beta_o \left[ - \frac{ 16\pi \Lambda^2 r^3_{co} }{9} + \frac{1-k}{2} \left(- \frac{8k\pi(k+\Lambda r^2_{\rm crit})}{r_{\rm crit}} + \frac{ 16\pi\Lambda GM_{\rm crit}}{3} \right) \right]   \,,
\end{eqnarray}
and the boundary term for the GB term gives
\begin{equation}
I^{\rm TE}_E({\rm GBb}) = x\alpha\beta_o 16\pi \left[\frac{\Lambda^2r^3_{co}}{9} - \frac{1-k}{2} \left( \frac{2G^2M^2_{\rm crit}}{r^3_{co}} + \frac{\Lambda GM_{\rm crit}}{3} \right) \right] \,.
\end{equation}
where we take $r_{co} \rightarrow \infty$. This gives $I^{\rm TEGB}_E(\rm total)$ to be zero for $k=1$ and the constant value for $k=-1$. However, the additional contribution of the free energy goes to zero, because of $r_{\rm crit}=\frac{1}{\sqrt{\Lambda}}$ for the extremal black hole. Therefore, the free energy turns out to be
\begin{equation}
F^{EGB}=  x\left( \frac{r_h(k-\frac{\Lambda}{3}r^2_h)}{8G} - \frac{k\alpha\kappa(k+\Lambda r^2_h)}{G r_h} + \frac{(1-k)}{12G\sqrt{\Lambda}} \right) \,.
\end{equation}
We mention about the reference background geometry in EGB theory. For $k=1$, the reference geometry does not have the horizon. As a result, the action, mass, entropy, and the free energy are all equal to zero. For $k=-1$, the entropy has a constant term due to the nonzero action with the constant term. However, the free energy does not have that contribution. As a result, there is no dependency of $\alpha$ for the reference background geometry. 	

\begin{figure}[t]
\begin{center}
\subfigure[For $k=1$ as a function of $r_h$]
{\includegraphics[width =2.3 in]{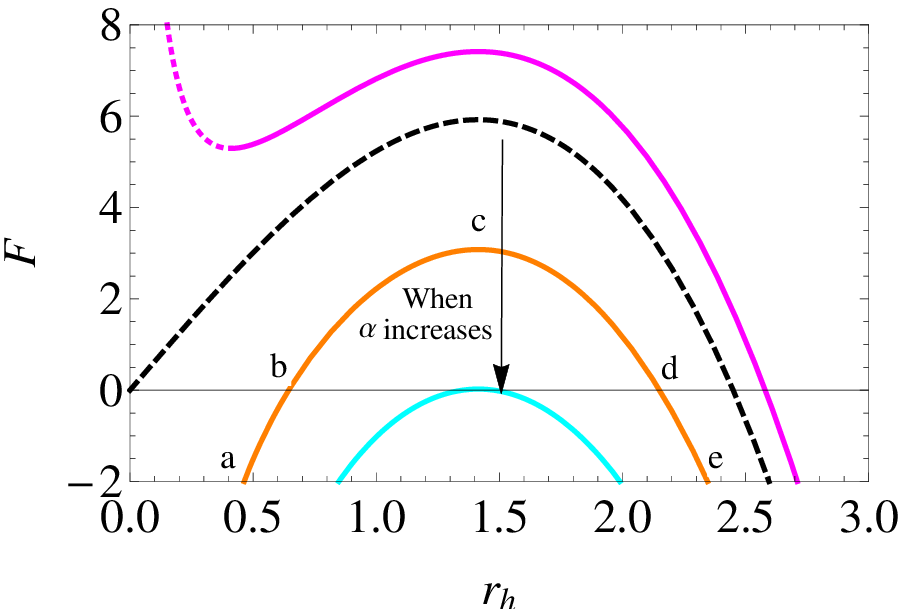}}
\subfigure[For $k=1$ as a function of $T_H$]
{\includegraphics[width =2.3 in]{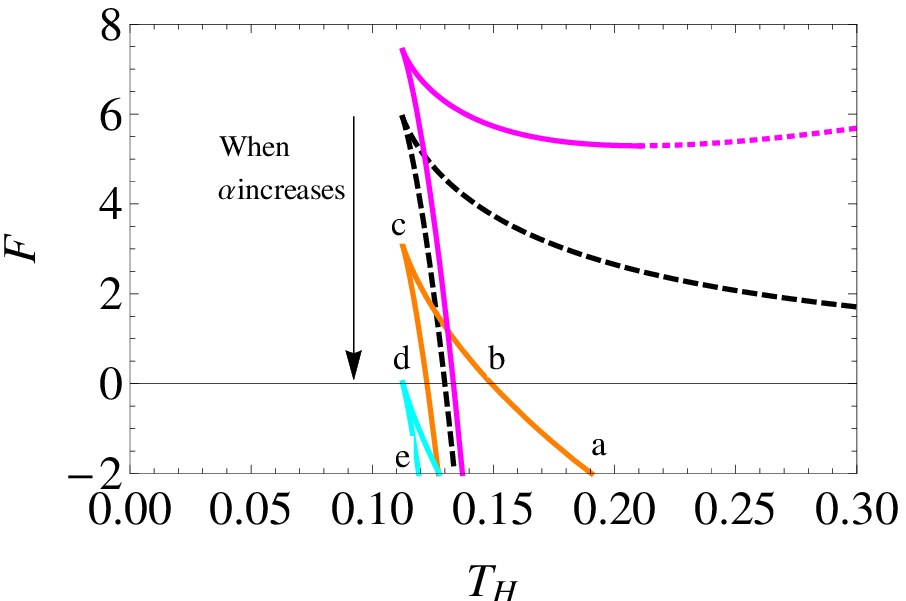}}\\
\subfigure[For $k=-1$ as a function of $r_h$]
{\includegraphics[width =2.3 in]{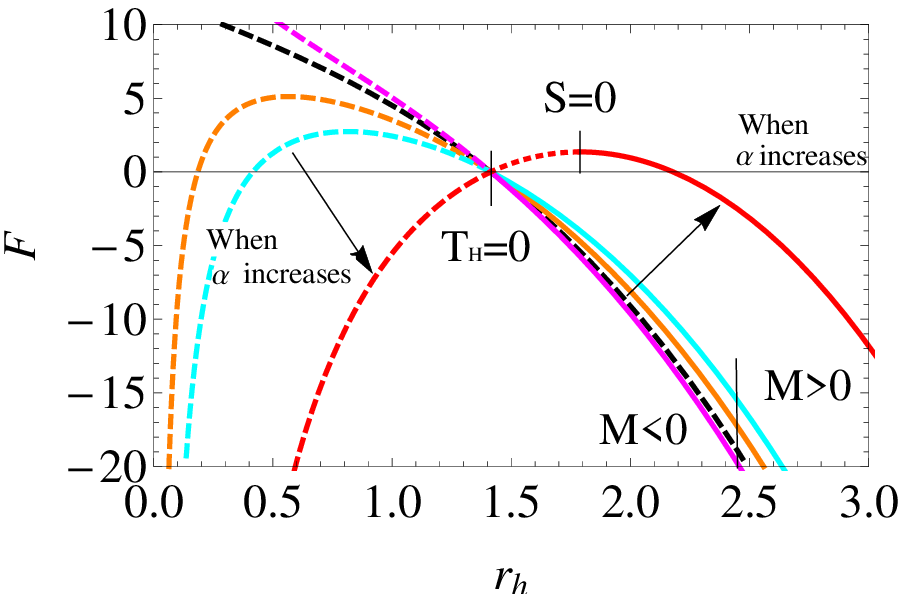}}
\subfigure[For $k=-1$ as a function of $T_H$]
{\includegraphics[width =2.3 in]{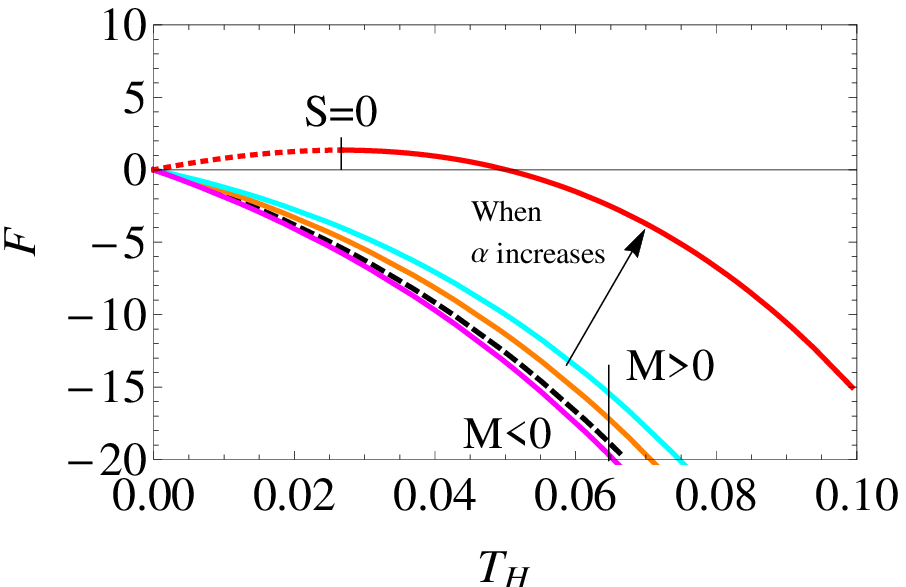}}
\end{center}
\caption{\footnotesize{(color online). Free energy difference in Einstein gravity and EGB theory both for $k=1$ and for $k=-1$ with $\Lambda=1/2$, and $\kappa=1$. $\alpha=0$ for the dashed line in Einstein theory. $\alpha=0.040$ for the orange line, $\alpha=0.083$ for the cyan line, $\alpha=0.800$ for the green line, and $\alpha=-0.021$ for the magenta line in EGB theory.}}
\label{free-energy-Einstein-k1}
\end{figure}

Figure \ref{free-energy-Einstein-k1} shows the behavior of the free energy difference between the black hole and the thermal  AdS with $k=1$ and $k=-1$ both in Einstein and in EGB theory. We choose the same parameters used for Fig.\ \ref{Entropy-EGB}. And again, we choose $\Lambda=1/2$ and $\kappa=1$. The black dashed line corresponds to $\alpha=0.000$ representing the free energy difference in Einstein theory. The orange line corresponds to $\alpha=0.040$. The cyan line to $\alpha=0.083$. The magenta line to $\alpha=-0.021$ with the cutoff at $r_h > \sqrt{0.168}$ representing the limitation of $r_h$ due to the positiveness of the entropy. Figure \ref{free-energy-Einstein-k1}(a) shows the behavior of the free energy difference as a function of the horizon radius $r_h$ for the positive mass case. The black dashed line with $\alpha=0.000$ has two regions, positive and negative parts. This one at $F=0$ indicates just the HP phase transition. If the value of $\alpha$ is increased for $0< \alpha < \frac{r^2_h(1-\frac{\Lambda}{3}r^2_h)}{8\kappa(1+\Lambda r^2_h)}$, then there exist three regions. The particular point is that there are stable small black holes belonging to the negative difference region, $r_h < b$ in the orange line, unlike those in Einstein theory. The black holes belonging to the positive difference region, $b < r_h < d$, are unstable like those in Einstein theory. The phase transition occurs at $r_h=b$ and $r_h=d$ in the orange line. When $\alpha$ increases, the line moves downward. If the value of $\alpha$ is larger than $\frac{1}{24\kappa\Lambda}$ the black holes are always stable more than the the thermal AdS space, which means that there is no phase transition. Figure \ref{free-energy-Einstein-k1}(b) shows the behavior of the difference as a function of the temperature $T_H$ using the relation $r_h=\frac{2\pi T_H \pm \sqrt{4\pi^2 T^2_H -k\Lambda}}{\Lambda}$. The left peaks at $T_H =c$ indicate the minimum temperature. The phase transition occurs from the thermal AdS space to the small black hole at the large critical temperature, $T_H=b$ in the orange line, unlike the case in Einstein theory. Over the large critical temperature, $b< T_H < a$, the small black hole can be more probable than the thermal AdS space.

\begin{figure}[t]
\begin{center}
\subfigure[For $k=1$ as a function of $r_h$]
{\includegraphics[width =2.3 in]{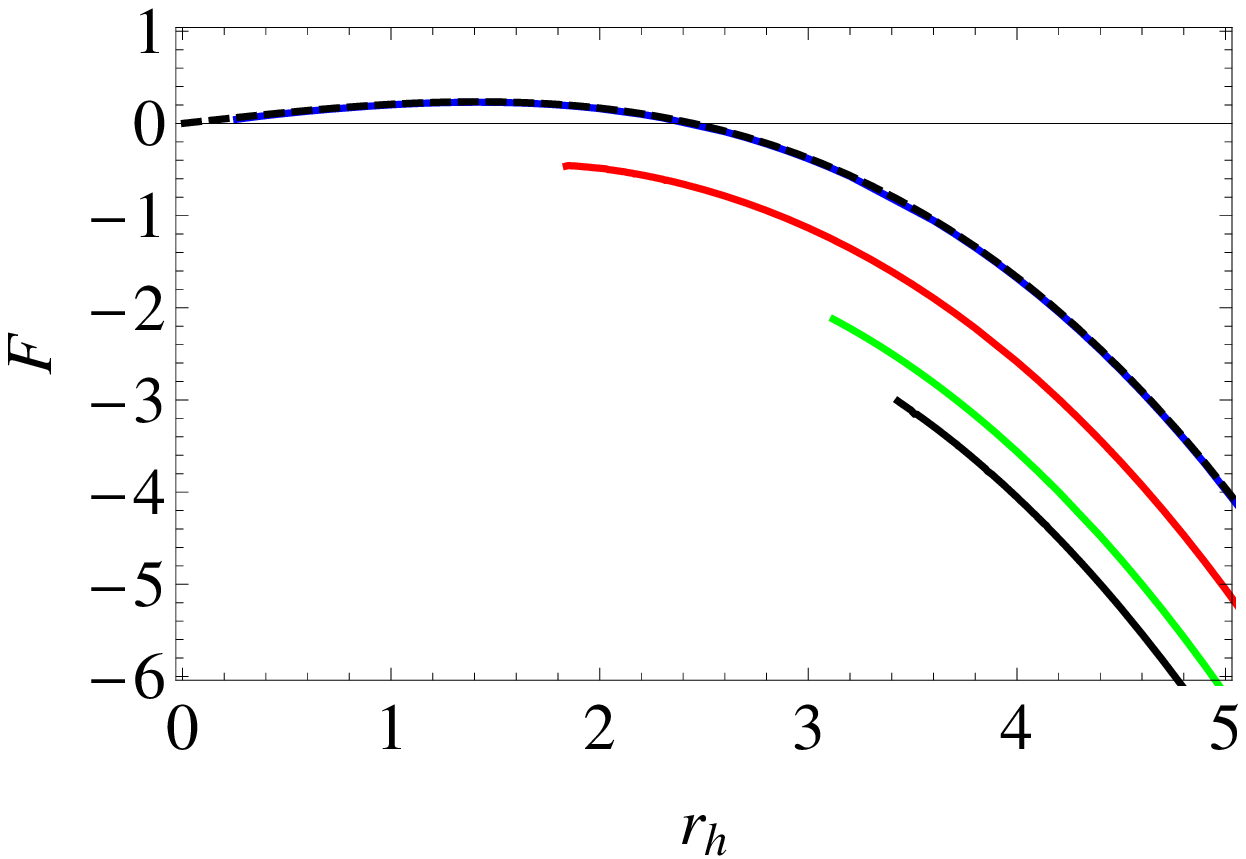}}
\subfigure[For $k=1$ as a function of $T_H$]
{\includegraphics[width =2.3 in]{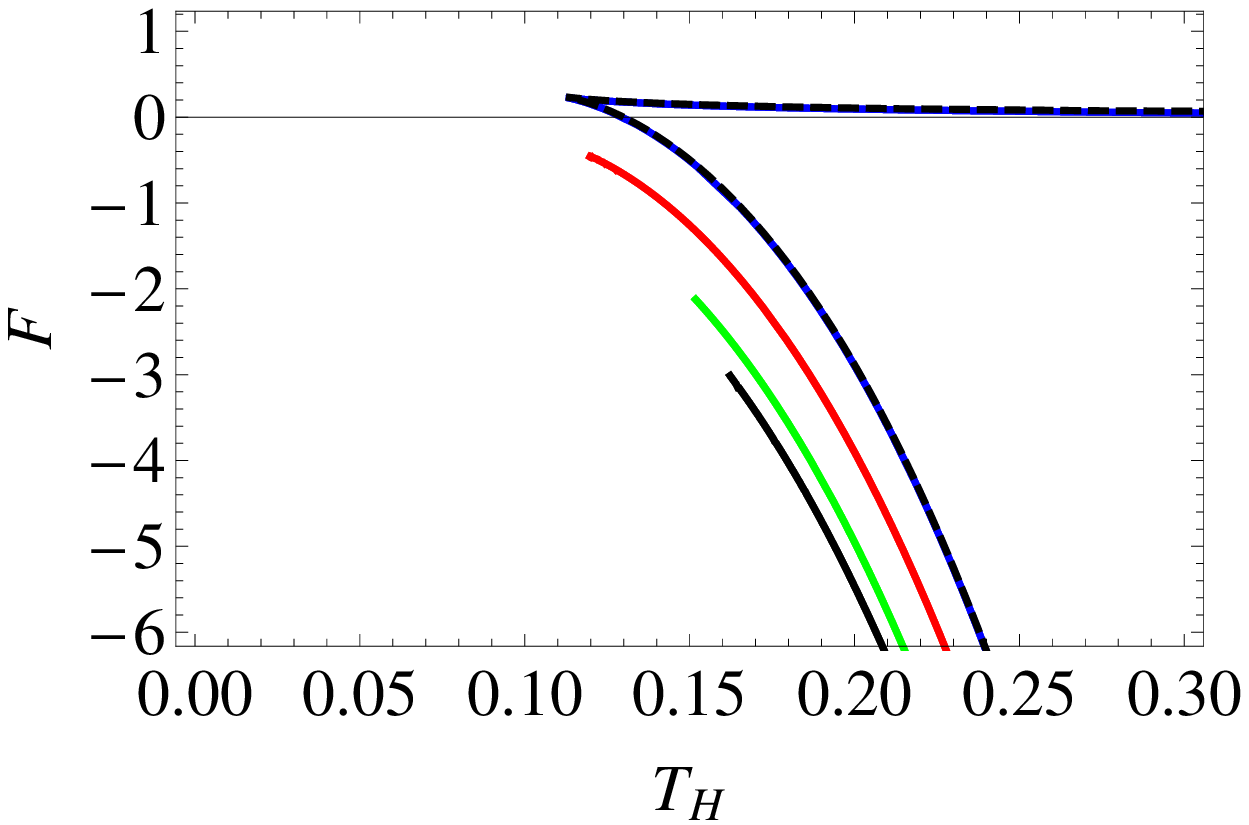}} \\
\subfigure[For $k=-1$ as a function of $r_h$]
{\includegraphics[width =2.3 in]{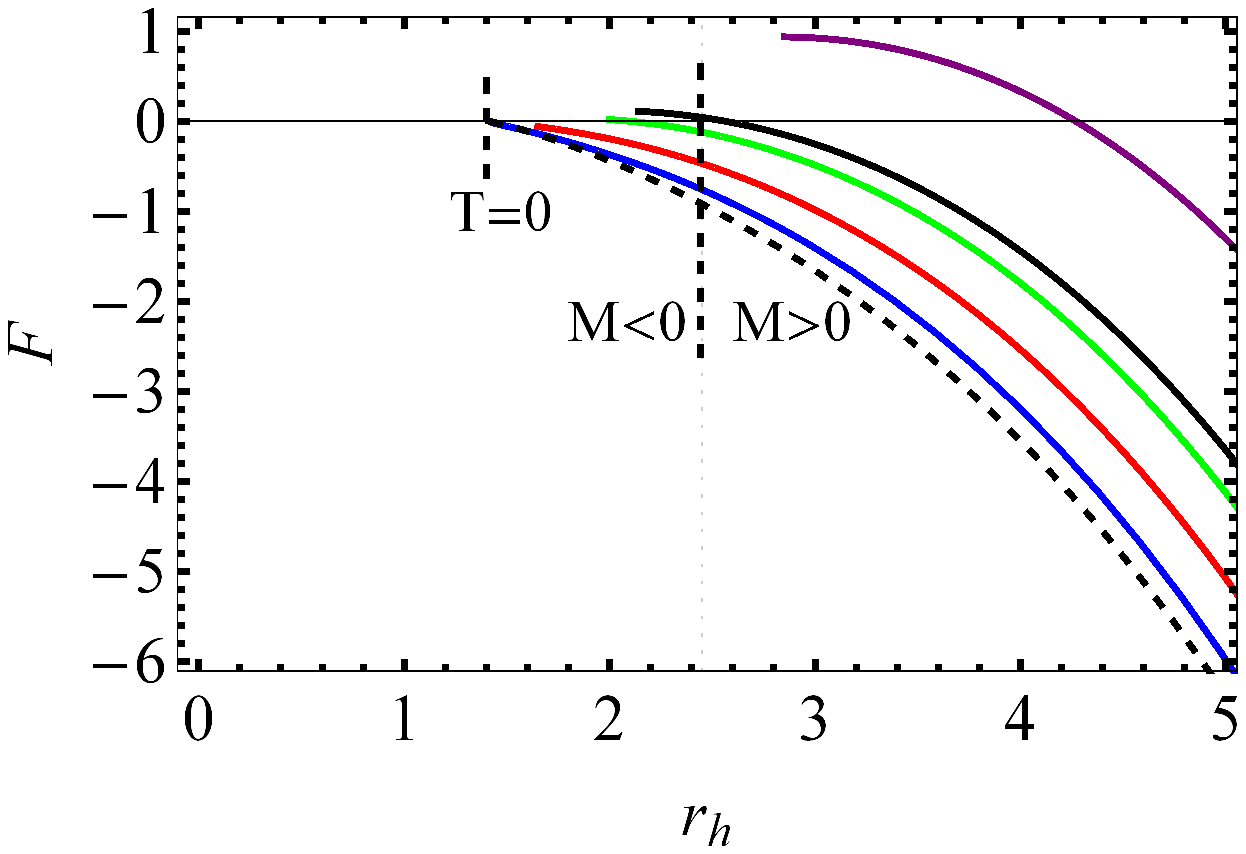}}
\subfigure[For $k=-1$ as a function of $T_H$]
{\includegraphics[width =2.3 in]{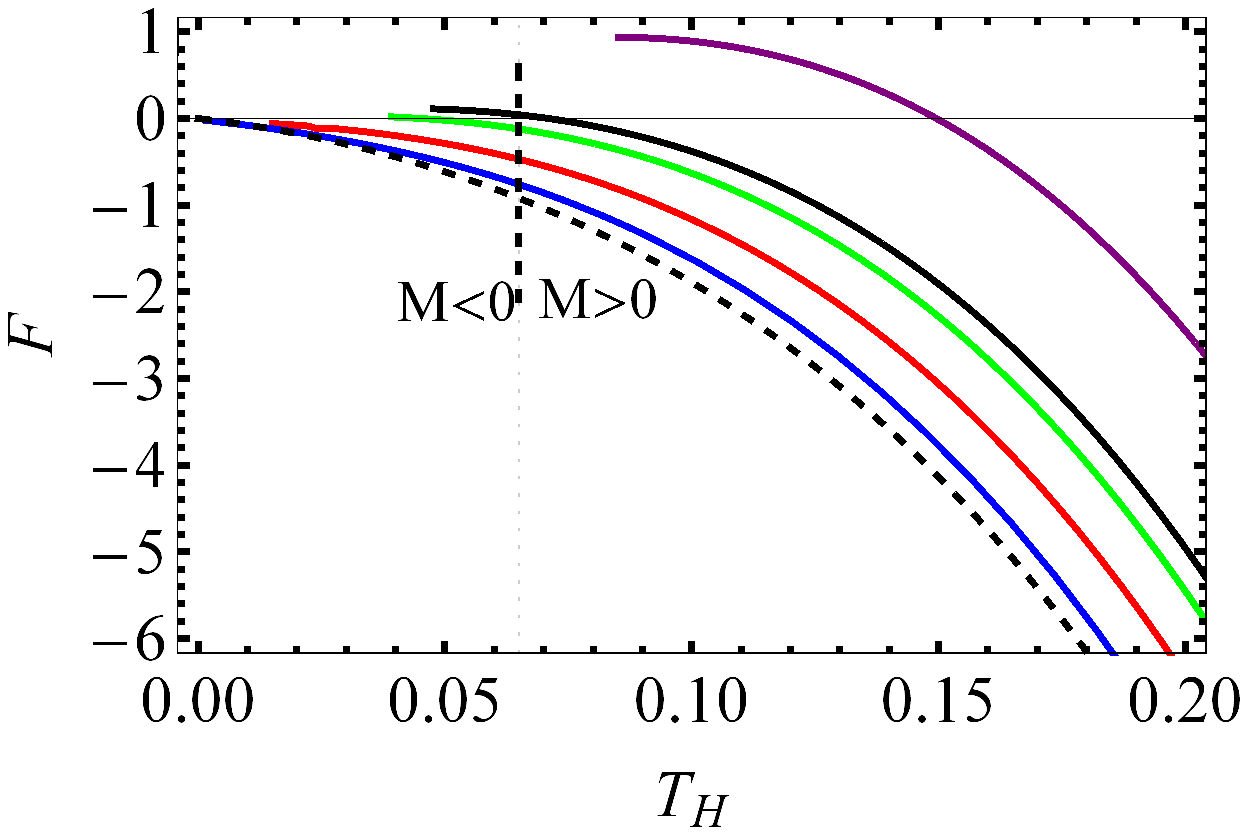}}
\end{center}
\caption{\footnotesize{(color online). The free energy difference both for $k=1$ and $k=-1$ with $\gamma=1/2$, $\Lambda=1/2$, and $\kappa=1$. $\alpha=0.000$ for the dashed line in Einstein theory. $\alpha=0.005$ for the blue line, $\alpha=0.400$ for the red line, $\alpha=0.800$ for the green line, $\alpha=1.000$ for the black line, and $\alpha=2.000$ for the purple line in DEGB theory.}}
\label{Free-energy-DEGB}
\end{figure}

For the case with $k=-1$, there are limitations for $r_h$ as shown in Fig.\ \ref{free-energy-Einstein-k1}(c). The dashed lines indicate those limitations. Thus we should erase the numerical data. The vertical line indicates the value with $M=0$. The left parts correspond to the case with the negative mass, while the right parts correspond to the case with the positive mass. The other limitation is from the positiveness of the entropy, $\alpha \leq \frac{r^2_h}{8 \kappa}$. The black line corresponds to the case of $\alpha=0.000$ in Einstein theory, in which there is no HP phase transition. When $\alpha$ increases, the line moves upward. If the value of $\alpha$ is greater than  $\frac{3\sqrt{\Lambda}r^2_h(1+\frac{\Lambda}{3}r^2_h)-4r_h}{24\kappa\sqrt{\Lambda}(-1+\Lambda r^2_h)}$, then the free energy difference has both the positive and the negative value to guarantee the phase transition. The red line showing the existence of the phase transition corresponds to the case with $\alpha=0.400$. The HP phase transition occurs from the thermal AdS space (or the extremal black hole) to the black hole.

Figure \ref{Free-energy-DEGB} shows the behavior of the free energy difference in DEGB theory. We employ the reference background geometry as those used for EGB theory. We use the thermodynamic relation $F=M-T_H S$ and feed numerical data into the relation. We choose the same parameters used for Fig.\ \ref{Entropy-DEGB}. And again, we take $\Lambda=1/2$ and $\gamma=1/2$ for four lines with $k=1$ and for five lines with $k=-1$. The dashed black line corresponds to $\alpha=0.000$ and $\lambda=0$ in Einstein theory, the blue line to $\alpha=0.005$ and $\lambda=1/600$, the red line to $\alpha=0.400$ and $\lambda=2/15$, the green line to $\alpha=0.800$ and $\lambda=4/15$, the black line to $\alpha=1.000$ and $\lambda=1/3$, and the purple line to $\alpha=2.000$ and $\lambda=2/3$. Figure \ref{Free-energy-DEGB}(a) shows the behavior of the free energy difference as a function of the horizon radius $r_h$. The blue line shows the HP phase transition occurs from the thermal AdS to the black hole geometry. When $\alpha$ increases, the line moves downward under the horizontal line of $F=0$. Consequentially, there is no phase transition. Figure \ref{Free-energy-DEGB}(b) shows the behavior of the difference as a function of the temperature $T_H$. For $k=-1$, the line moves upward over the horizontal line of $F=0$ when $\alpha$ increases. The black and purple lines have both the positive and the negative free energy difference to guarantee the phase transition.

\section{SUMMARY AND DISCUSSION \label{sec4}}

We have studied the black hole thermodynamics and Hawking-Page phase transition in asymptotically AdS spacetime in the (dilatonic) Einstein-Gauss-Bonnet theory of gravitation. We considered both the spherical for $k=1$ and the hyperbolic black holes for $k=-1$.

First, we have constructed the hairy black holes by solving the coupled equations of motion for the gravity and the scalar field simultaneously. We have followed the procedure to be consistent with our action according to Refs.\ \cite{got04, ohto04}. If the black hole horizon becomes larger, the magnitude of the scalar hair becomes smaller. There exists the minimum mass of a black hole under the given parameter sets.

Second, we have calculated the temperature and the specific heat for the AdS black hole solutions. If the value of $\alpha$ becomes larger, the minimum mass of the black hole becomes larger. For $k=1$, if the value of $\alpha$ becomes positively larger, there exist the black hole only belonging to the increasing part of the temperature, which corresponds to the positive specific heat unlike the AdS black hole both in Einstein theory and in the EGB theory. For $k=-1$, the black holes belong to the increasing part of the temperature, which corresponds to the positive specific heat similarly to the AdS black hole in the theory with the vanishing scalar field. When the specific heat has the positive value, one can employ the canonical ensemble description.

We have divided the calculations into two aspects for the rest of the investigation, either EGB theory with analytical calculations or DEGB theory with numerical computations. Third, we have calculated straightforwardly the Euclidean action of a black hole in AdS spacetime in EGB theory. We calculated the entropy using the thermodynamic relation. There is the additional constant term depending on the values of $\alpha$ and $k$ in the entropy formula. That is why we did not consider the black holes with $k=0$. From the positiveness of the entropy, the values are restricted by the signs of both $\alpha$ and $k$. We have adopted the known entropy formula for the DEGB theory and fed the numerical data into the formula.

Last, we have calculated the free energy difference to analyze the HP phase transition in (D)EGB theories. We calculated the free energy using the Euclidean action in EGB theory and using the thermodynamic relation in DEGB theory. For $k=1$, we take the reference background geometry as the thermal AdS space with $M=0$, in which the Euclidean action for this space vanishes. Therefore, the pure quantity of the black hole free energy becomes important. We have obtained the qualitatively different behavior of the free energy difference depending on the parameters. The free energy difference decreases as the value of $\alpha$ positively increases. In other words, the phase transition occurs from the thermal AdS space to the small black hole unlike that in Einstein theory. Hence, there exist globally stable small black holes if the value of $\alpha$ increases. Consequently, black holes are definitively divided into four types in terms of the radius of a black hole. The new one belongs to locally unstable and globally stable phase. Because of the local instability by the negative specific heat, the black hole evaporates. If $\alpha > \frac{1}{24\kappa\Lambda}$, all black holes belong to globally stable phase. Therefore, there is no phase transition unlike that in Einstein theory. For $k=-1$, we take the reference background geometry as the thermal AdS space with $M=M_{\rm crit}$, in which the geometry with $M_{\rm crit}$ corresponds to the extremal black hole with the negative mass. If the value of $\alpha$ increases, the HP phase transition occurs in (D)EGB theory unlike that in Einstein theory.

We now discuss the reference background geometry as the qualitatively different one. The extremal charged black hole solution has the vanishing entropy. Because the extremal black hole can be in thermodynamic equilibrium with the thermal radiation at any temperature, one can introduce an arbitrary temperature in this geometry \cite{hhro, cteit}. For $k=1$, the thermal AdS spacetime with $M=0$ corresponds to the reference geometry without a black hole. This geometry has the vanishing action, mass, entropy, and free energy both for Einstein and for EGB theory. For $k=-1$, the thermal AdS with $M=M_{\rm crit}$ corresponds to the reference geometry with the extremal hyperbolic black hole. This geometry has the vanishing entropy. Therefore, we have taken the extremal black hole geometry as the reference one and introduced the arbitrary temperature in Einstein theory. The entropy has a constant term due to the nonzero action with the constant term in EGB theory. However, the free energy does not have the contribution. We emphasize that we have calculated the Euclidean action, entropy, and free energy of the reference background geometry both in Einstein theory and in EGB theory. They do not correspond to the simple limiting cases of the nonzero mass black hole and nonextremal black hole geometry. There exist discontinuous jumps in those physical quantities depending on the GB term. For the DEGB theory, we have employed the reference geometry as those used for EGB theory. If one could find the soliton solution with the dilaton hair in the asymptotically AdS spacetime in DEGB theory, the geometry of the soliton solution could be the reference geometry. We expect that the resulting free energy difference is not qualitatively different from those in our analysis, because the free energy for that reference geometry is not expected to be significantly changed.

We note that the relation, $F=M-T_H S$, in this paper was used for the free energy, in which the common redshift factor in both sides were eliminated. The conversed mass $M$ is the thermodynamic internal energy $E$ divided by the redshift factor, which reduces to the mass of the black hole at spatial infinity \cite{bcma}. According to the above relation, the total amount of energy $E$ is composed of two energies, the free energy and the product of the entropy and the temperature. The free energy can be transformed into work, i.e.\ this energy is useful, while the other energy given by the product cannot be transformed into work, i.e.\ this energy is useless. If the entropy at the fixed temperature is increased by the GB term, the free energy is decreased by the same quantity at the fixed total energy, and vice versa. Hence the GB term seems to increase or decrease the useful energy depending on the signs of $\alpha$ and $k$ with the fixed total energy.

We note that the thermodynamic instabilities are divided into two broad categories, either global instability determined by the sign of the free energy difference in the grand canonical ensemble or local instability determined by the sign of the specific heat (or heat capacity) in the canonical ensemble \cite{rgc0, ssse, ysm11}. There is no electromagnetic charge in our system. Therefore, the grand canonical ensemble reduces to the canonical ensemble. The local stability of the thermodynamic system means that small perturbations do not remove the system from its equilibrium \cite{mjdo} similarly to the particle in the potential well. While, the global stability of the system means that the system is in the stable phase similarly to the lowest vacuum state in the asymmetric double well potential, in which the HP phase transition occurs between two phases at the same temperature or the zero free energy difference \cite{kolp}.

In this paper we have shown that the GB term influences the entropy and the free energy without modifying the temperature and the specific heat, while the GB term with the dilaton coupling influences all physical quantities. In other words, the GB term affects the phase transition without modifying the local thermodynamic instability, while if the GB term is coupled to the scalar field both global and local instabilities are affected.

In four dimensions, the existence of the GB term does not influence both the equations of motion and the solutions. The observer cannot distinguish the difference between the Einstein theory and EGB theory. However, the observer can distinguish the differences of thermodynamic properties and phase transition between two theories. For this reason, the observation of the HP phase transition can determine which theory of gravitation describes our Universe. In other words, we think that the topological information of the spacetime manifold could be additionally stored in the thermodynamic quantities for a black hole.

The HP phase transition is the first-order phase transition, because the transition has a discontinuous jump in the first-order derivatives of the free energy difference. In this paper, we have considered the thermal AdS spacetime, the AdS black hole without hair, and the black hole with hair. We have obtained the free energy showing the phase transition between the thermal AdS and the AdS black hole geometry. It will be interesting to investigate whether the phase transition occurs between the black hole (with/without hair) and the other black hole (with/without hair) corresponding to the reference geometry, and is the first-order or the second-order phase transition. It will be also interesting to investigate whether or not the negative mode exists. The black hole solutions in AdS spacetime could be important for holographic applications of strongly coupled field theories. Recently, there have been a wide variety of developments for holographic superconductors, critical behaviors, and hyperscaling violation \cite{Hartnoll:2008vx, Charmousis:2010zz, Gouteraux:2011ce, Cadoni:2012uf, Bai:2014poa, Cai:2015cya}. In this paper, we have restricted our study to asymptotically Schwarzschild-AdS solutions. However, solutions with different asymptotic, for instance, hyperscaling violation in modified spacetime, could be investigated. These kinds of geometries could draw interest for both phase transition phenomena and holography applications, which have been considered extensively in Refs.\ \cite{Gouteraux:2011ce, Cadoni:2012uf} and references therein. The DEGB theory could bring a possibility of such phase transition phenomena as well as holography application. We postpone any possible application of such phase transitions and holographic applications including the meaning of the order in the phase transition for our future work.

\section*{ACKNOWLEDGEMENTS}
We would like to thank Yun Soo Myung, Wontae Kim, Mu-In Park, Hyun Seok Yang, Chanyong Park, Seoktae Koh, and Eugen Radu for helpful discussions and comments. We would like to thank Hyung Won Lee, Kyoung Yee Kim, and Jeongcho Kim for their hospitality during our visit to Inje University. B.H.L. was supported by the National Research Foundation of Korea(NRF) grant funded by the Korea government(MSIP) (No. 2014R1A2A1A01002306). W.L. was supported by Basic Science Research Program through the National Research Foundation of Korea(NRF) funded by the Ministry of Education, Science and Technology(2016R1D1A1B01010234).


\newpage

\begin{thebibliography}{99}

\bibitem{beken} J.~D.~Bekenstein,
    Phys.\ Rev.\ D {\bf 7}, 2333 (1973).
\bibitem{hawen} S.~W.~Hawking,
    Commun.\ Math.\ Phys.\ {\bf 43}, 199 (1975);
    Erratum: [Commun.\ Math.\ Phys.\  {\bf 46}, 206 (1976)].
\bibitem{jamy} T.~Jacobson and R.~C.~Myers,
    Phys.\ Rev.\ Lett.\ {\bf 70}, 3684 (1993)
    [hep-th/9305016].
\bibitem{wald8} R. M. Wald,
    Phys.\ Rev.\ D {\bf 48}, R3427 (1993)
    [gr-qc/9307038].
\bibitem{giha} G.~W.~Gibbons and S.~W.~Hawking,
    Phys.\ Rev.\ D {\bf 15}, 2752 (1977).
\bibitem{bcha} J.~M.~Bardeen, B.~Carter, and S.~W.~Hawking,
    Commun.\ Math.\ Phys.\ {\bf 31}, 161 (1973).
\bibitem{DeWitt:1975ys}
    B.~S.~DeWitt,
    Phys.\ Rept.\  {\bf 19}, 295 (1975).
\bibitem{swh} S.~W.~Hawking,
    Nucl.\ Phys.\ B {\bf 144}, 349 (1978).
\bibitem{tkm3} T.~Jacobson, G.~Kang, and R.~C.~Myers,
    Phys.\ Rev.\ D {\bf 49}, 6587 (1994)
    [gr-qc/9312023].
\bibitem{Liko:2007vi}
    T.~Liko,
    Phys.\ Rev.\ D {\bf 77}, 064004 (2008)
    [arXiv:0705.1518 [gr-qc]].
\bibitem{sawa} S.~Sarkar and A.~C.~Wall,
    Phys.\ Rev.\ D {\bf83}, 124048 (2011)
    [arXiv:1011.4988 [gr-qc]].
\bibitem{HawPa} S.~W.~Hawking and D.~N.~Page,
    Commun.\ Math.\ Phys.\ {\bf 87}, 577 (1983).
\bibitem{ysm10} Y.~S.~Myung,
    Phys.\ Lett.\ B {\bf 638}, 515 (2006)
    [gr-qc/0603051].
\bibitem{ekyi} M.~Eune, W.~Kim, and S.-H.~Yi,
    J.\ High Energy Phys.\ 03 (2013) 020
    [arXiv:1301.0395 [gr-qc]].
\bibitem{mymo} Y.~S.~Myung and T.~Moon,
    J.\ High Energy Phys.\ 04 (2014) 058
    [arXiv:1311.6985 [hep-th]].
\bibitem{giki} Y.~Gim and W.~Kim,
    J.\ Cosmol.\ Astropart.\ Phys.\ 10 (2014) 003
    [arXiv:1406.6475 [gr-qc]].
\bibitem{ysm11} Y.~S.~Myung,
    Adv.\ High Energy Phys.\ {\bf 2015}, 478273 (2015)
    [arXiv:1510.02853 [gr-qc]].
\bibitem{egki} M.~Eune, Y.~Gim, and W.~Kim,
    Phys.\ Rev.\ D {\bf 91}, 044037 (2015)
    [arXiv:1409.5548 [gr-qc]].
\bibitem{dema}
    A.~Dey, S.~Mahapatra, and T.~Sarkar,
    Phys.\ Rev.\ D {\bf 94}, 026006 (2016)
    [arXiv:1512.07117 [hep-th]].
\bibitem{maha} S.~Mahapatra,
    J.\ High Energy Phys.\ 04 (2016) 142
    [arXiv:1602.03007 [hep-th]].
\bibitem{rgc0} R.-G.~Cai,
    Phys.\ Rev.\ D {\bf65}, 084014 (2002)
    [hep-th/0109133].
\bibitem{chne} Y.~M.~Cho and I.~P.~Neupane,
    Phys.\ Rev.\ D {\bf 66}, 024044 (2002)
    [hep-th/0202140].
\bibitem{neup} I.~P.~Neupane,
    Phys.\ Rev.\ D {\bf 69}, 084011 (2004)
    [hep-th/0302132].
\bibitem{cakw} R.-G.~Cai, S.~P.~Kim, and B.~Wang,
    Phys.\ Rev.\ D {\bf 76}, 024011 (2007)
    [arXiv:0705.2469 [hep-th]].
\bibitem{mkp}  Y.~S.~Myung, Y.-W.~Kim, and Y.-J. Park,
    Eur.\ Phys.\ J.\ C {\bf58}, 337 (2008)
    [arXiv:0806.4452 [gr-qc]].
\bibitem{campf} C.~G.~Callan, Jr., E.~J.~Martinec, M.~J.~Perry, and D.~Friedan,
    Nucl.\ Phys.\ B {\bf 262}, 593 (1985).
\bibitem{Zwiebach} B.~Zwiebach,
    Phys.\ Lett.\ B {\bf 156}, 315 (1985).
\bibitem{boude} D.~G.~Boulware and S.~Deser,
    Phys.\ Lett.\ B {\bf 175}, 409 (1986).
\bibitem{anrt} M.~Gasperini, M.~Maggiore, and G. Veneziano,
    Nucl.\ Phys.\ B {\bf 494}, 315 (1997)
    [hep-th/9611039].
\bibitem{kllt} S.~Koh, B.-H.~Lee, W.~Lee, and G.~Tumurtushaa,
    Phys.\ Rev.\ D {\bf 90}, 063527 (2014)
    [arXiv:1404.6096 [gr-qc]].
\bibitem{agll} W.-K.~Ahn, B.~Gwak, B.-H.~Lee, and W.~Lee,
    Eur.\ Phys.\ C {\bf 75}, 372 (2015)
    [arXiv:1412.4189 [gr-qc]].
\bibitem{copw} S.~R.~Coleman, J.~Preskill, and F.~Wilczek,
    Nucl.\ Phys.\ B {\bf 378}, 175 (1992)
    [hep-th/9201059].
\bibitem{kanti01} P.~Kanti, N.~E.~Mavromatos, J.~Rizos, K.~Tamvakis, and E.~Winstanley,
    Phys.\ Rev.\ D {\bf 54}, 5049 (1996)
    [hep-th/9511071].
\bibitem{tyma02} T.~Torii, H.~Yajima, and K.-i.~Maeda,
    Phys.\ Rev.\ D {\bf 55}, 739 (1997)
    [gr-qc/9606034].
\bibitem{kanti02} P.~Kanti, N.~E.~Mavromatos, J.~Rizos, K.~Tamvakis, and E.~Winstanley,
    Phys.\ Rev.\ D {\bf 57}, 6255 (1998)
    [hep-th/9703192].
\bibitem{toma3} T.~Torii and K.-i.~Maeda,
    Phys.\ Rev.\ D {\bf 58}, 084004 (1998).
\bibitem{got03} Z.~K.~Guo, N.~Ohta, and T.~Torii,
    Prog.\ Theor.\ Phys.\ {\bf 120}, 581 (2008)
    [arXiv:0806.2481 [gr-qc]].
\bibitem{got04} Z.~K.~Guo, N.~Ohta, and T.~Torii,
    Prog.\ Theor.\ Phys.\  {\bf 121}, 253 (2009)
    [arXiv:0811.3068 [gr-qc]].
\bibitem{ohto04} N.~Ohta and T.~Torii,
    Prog.\ Theor.\ Phys.\  {\bf 121}, 959 (2009)
    [arXiv:0902.4072 [hep-th]].
\bibitem{tkm4} T.~Jacobson, G.~Kang, and R.~C.~Myers,
    Phys.\ Rev.\ D {\bf 52}, 3518 (1995)
    [gr-qc/9503020].
\bibitem{bakr} V.~Balasubramanian and P.~Kraus,
    Commun.\ Math.\ Phys.\ {\bf 208}, 413 (1999),
    [hep-th/9902121].
\bibitem{ejmy} R.~Emparan, C.~V.~Johnson, and R.~C.~Myers,
    Phys.\ Rev.\ D {\bf 60}, 104001 (1999)
    [hep-th/9903238].
\bibitem{lisa} J.~T.~Liu and W.~A.~Sabra,
    Class.\ Quantum Grav.\ {\bf 27}, 175014 (2010)
    [arXiv:0807.1256 [hep-th]].
\bibitem{park01} C.~Park,
    Adv.\ High Energy Phys.\ {\bf 2013}, 389541 (2013)
    [arXiv:1209.0842 [hep-th]].
\bibitem{park02} C.~Park,
    Phys.\ Rev.\  D {\bf89}, 066003 (2014)
    [arXiv:1312.0826 [hep-th]].
\bibitem{York} J.~W.~York, Jr.,
    Phys.\ Rev.\ Lett.\ {\bf 28}, 1082 (1972).
\bibitem{Myers:1987yn}
    R.~C.~Myers,
    Phys.\ Rev.\ D {\bf 36}, 392 (1987).
\bibitem{Davis:2002gn}
    S.~C.~Davis,
    Phys.\ Rev.\ D {\bf 67}, 024030 (2003)
    [hep-th/0208205].
\bibitem{Brihaye:2008xu}
    Y.~Brihaye and E.~Radu,
    J.\ High Energy Phys.\ 09 (2008) 006
    [arXiv:0806.1396 [gr-qc]].
\bibitem{misner} C.~W.~Misner, K.~S.~Thorne, and J.~A.~Wheeler, {\em Gravitation} (W.~H.~Freeman and Company, New York, 1973).
\bibitem{PerezBergliaffa:2011gj}
    S.~E.~Perez Bergliaffa and Y.~E.~C.~de Oliveira Nunes,
    Phys.\ Rev.\ D {\bf 84}, 084006 (2011)
    [arXiv:1107.5727 [gr-qc]].
\bibitem{bcma} J.~D.~Brown, J.~Creighton, and R.~B.~Mann,
    Phys.\ Rev.\ D {\bf 50}, 6394 (1994)
    [gr-qc/9405007].
\bibitem{robm} R.~Emparan,
    Phys.\ Lett.\ B {\bf 432}, 74 (1998)
    [hep-th/9804031].
\bibitem{birm} D.~Birmingham,
    Classical Quantum Gravity  {\bf 16}, 1197 (1999)
    [hep-th/9808032].
\bibitem{sugo} D.~Sudarsky and J.~A.~Gonzalez,
    Phys.\ Rev.\ D {\bf 67}, 024038 (2003)
    [gr-qc/0207069].
\bibitem{agpa} C.~Arg\"{u}elles, N.~Grandi, and M.-I.~Park,
    J.\ High Energy Phys.\ 10 (2015) 100
    [arXiv:1508.04380 [hep-th]].
\bibitem{vanzo} L.~Vanzo,
    Phys.\ Rev.\ D {\bf56}, 6475 (1997)
    [gr-qc/9705004].
\bibitem{cpark5} C.~Park,
    arXiv:1606.07340 [hep-th].
\bibitem{hhro} S.~W.~Hawking, G.~T.~Horowitz, and S.~F.~Ross,
    Phys.\ Rev.\ D {\bf 51}, 4302 (1995)
    [gr-qc/9409013].
\bibitem{cteit} C.~Teitelboim,
    Phys.\ Rev.\ D {\bf 51}, 4315 (1995);
    Erratum: [Phys.\ Rev.\ D {\bf 52}, 6201 (1995)]
    [hep-th/9410103].
\bibitem{ssse} A.~Sahay, T.~Sarkar, and G.~Sengupta,
    J.\ High Energy Phys.\ 04 (2010) 118
    [arXiv:1002.2538 [hep-th]].
\bibitem{mjdo} M.~J.~de Oliveira, {\em Equilibrium Thermodynamics} (Springer-Verlag, Berlin, 2013).
\bibitem{kolp} B.~Kol,
    Phys.\ Rept.\  {\bf 422}, 119 (2006)
    [hep-th/0411240].
\bibitem{Hartnoll:2008vx}
    S.~A.~Hartnoll, C.~P.~Herzog, and G.~T.~Horowitz,
    Phys.\ Rev.\ Lett.\  {\bf 101}, 031601 (2008)
    [arXiv:0803.3295 [hep-th]].
\bibitem{Charmousis:2010zz}
    C.~Charmousis, B.~Gouteraux, B.~S.~Kim, E.~Kiritsis, and R.~Meyer,
    J.\ High Energy Phys.\ 11 (2010) 151
    [arXiv:1005.4690 [hep-th]].
\bibitem{Gouteraux:2011ce}
    B.~Gouteraux and E.~Kiritsis,
    J.\ High Energy Phys.\ 12 (2011) 036
    [arXiv:1107.2116 [hep-th]].
\bibitem{Cadoni:2012uf}
    M.~Cadoni and S.~Mignemi,
    J.\ High Energy Phys.\ 06 (2012) 056
    [arXiv:1205.0412 [hep-th]].
\bibitem{Bai:2014poa}
    X.~Bai, B.~H.~Lee, M.~Park, and K.~Sunly,
    J.\ High Energy Phys.\ 09 (2014) 054
    [arXiv:1405.1806 [hep-th]].
\bibitem{Cai:2015cya}
    R.~G.~Cai, L.~Li, L.~F.~Li, and R.~Q.~Yang,
    Sci.\ China Phys.\ Mech.\ Astron.\  {\bf 58}, 060401 (2015)
    [arXiv:1502.00437 [hep-th]].















\end{thebibliography}
\end{document}